\documentclass[aps,jcp,preprint,superscriptaddress]{revtex4}
\usepackage{graphicx}
\usepackage{amsmath}
\usepackage{dcolumn}
\usepackage{bm}

\begin{document}
\title{On the role of the magnetic dipolar interaction in cold and ultracold collisions: Numerical and
analytical results for NH($^3\Sigma^-$)~+~NH($^3\Sigma^-$)}

\author{Liesbeth M.~C.~Janssen}
\affiliation{Radboud University Nijmegen, Institute for Molecules and Materials,
Heyendaalseweg 135, 6525 AJ Nijmegen, The Netherlands}
\author{Ad van der Avoird}
\affiliation{Radboud University Nijmegen, Institute for Molecules and Materials,
Heyendaalseweg 135, 6525 AJ Nijmegen, The Netherlands}
\author{Gerrit C.~Groenenboom}
\email[Electronic mail: ]{gerritg@theochem.ru.nl}
\affiliation{Radboud University Nijmegen, Institute for Molecules and Materials,
Heyendaalseweg 135, 6525 AJ Nijmegen, The Netherlands}

\date{\today}

\begin{abstract}
We present a detailed analysis of the role of the magnetic dipole-dipole
interaction in cold and ultracold collisions. We focus on collisions between
magnetically trapped NH molecules, but the theory is general for any two
paramagnetic species for which the electronic spin and its space-fixed projection
are (approximately) good quantum numbers. 
It is shown that dipolar spin relaxation is directly
associated with magnetic-dipole induced avoided crossings that occur between
different adiabatic potential curves. For a given collision energy and magnetic
field strength, the cross-section contributions from different scattering
channels depend strongly on whether or not the corresponding avoided
crossings are energetically accessible. We find that the crossings become lower
in energy as the magnetic field decreases, so that higher partial-wave
scattering becomes increasingly important \textit{below} a certain magnetic
field strength. 
In addition, we derive analytical cross-section expressions for
dipolar spin relaxation based on the Born approximation and distorted-wave Born
approximation. The validity regions of these analytical expressions are
determined by comparison with the NH + NH cross sections obtained from full coupled-channel
calculations. We find
that the Born approximation is accurate over a wide range of energies and
field strengths, but breaks down at high energies and high magnetic fields.
The analytical distorted-wave Born approximation gives more accurate results in the case of
$s$-wave scattering, but shows some significant discrepancies for the higher partial-wave
channels.
We thus conclude that the Born approximation gives generally more
meaningful results than the distorted-wave Born approximation at the collision
energies and fields considered in this work.

\end{abstract}

\maketitle

\section{Introduction}

The ability to produce and trap atomic and molecular species at sub-kelvin
temperatures offers numerous exciting possibilities in condensed-matter
physics \cite{greiner:02,jaksch:05,micheli:06,bloch:08}, quantum computing \cite{demille:02,gulde:03,andre:06},
high-precision spectroscopy \cite{lev:06,fortier:07,bethlem:09,tarbutt:09,poli:11}, and physical
chemistry \cite{meerakker:05,gilijamse:06,gilijamse:07,campbell:07,campbell:08,sawyer:08b,krems:08,
campbell:09,scharfenberg:10,ospelkaus:10}. 
Since the experimental realization of the first Bose-Einstein
condensates \cite{anderson:95a,davis:95}, major advances have been made in the field of ultracold atomic
gases. It is now well established that alkali-metal atoms can be efficiently
cooled into the ultracold regime using a combination of laser cooling and
evaporative cooling.  However, laser cooling is not applicable to all atomic
species, and is particularly difficult for molecules \cite{shuman:10}. In the last few years,
several methods have been developed that aim at 
producing (ultra)cold molecular gases at relatively high densities.
Techniques such
as photoassociation \cite{jones:06} and magnetic Feshbach association \cite{kohler:06} employ an indirect 
scheme in which ultracold molecules are formed by pairing up
pre-cooled, ultracold atoms. These methods are, however, currently limited to molecules
consisting of two alkali-metal atoms. Direct-cooling methods such as Stark \cite{bethlem:03} and Zeeman
deceleration \cite{narevicius:08}, molecular-beam guiding \cite{rieger:05},
and buffer-gas cooling \cite{weinstein:98} apply to a much
wider range of molecular species, but require a second-stage cooling technique 
to reach the ultracold regime.  Although several theoretical studies have shown
that e.g.\ sympathetic cooling of cold molecules with ultracold co-trapped
atoms \cite{soldan:09,wallis:09,wallis:10,barletta:09,barletta:10} or molecular evaporative
cooling \cite{avdeenkov:01,janssen:11,janssen:11a} is likely to be successful, 
this is yet to be demonstrated experimentally.

Second-stage cooling methods such as forced evaporative cooling require strong
elastic collisions that thermalize the gas cloud as the trap depth is slowly
reduced \cite{ketterle:96}. Inelastic collisions, in which the internal quantum
state of at least one of the collision partners is changed, can induce heating
of the gas and trap loss.  A detailed understanding of the interparticle
interactions that govern these inelastic processes is thus crucial for
assessing the feasibility of second-stage cooling. One of the most important
inelastic loss mechanisms for trapped paramagnetic species is dipolar spin
relaxation, which arises from the magnetic dipole-dipole interaction between
the magnetic moments of the particles. For many spin-polarized
atomic gases such as hydrogen \cite{ketterle:96}, lithium \cite{gerton:99},
nitrogen \cite{tscherbul:10}, and chromium \cite{hensler:03}, but also for
atom--molecule and molecule--molecule systems such as Li + NH \cite{wallis:10},
N + NH \cite{hummon:11}, and NH + NH \cite{janssen:11,janssen:11a}, the
interparticle dipolar spin-spin interaction is indeed the dominant source of trap loss.

In this paper, we provide a comprehensive study on the role of the magnetic dipolar
interaction in cold and ultracold collisions. Specifically, we consider collisions
between magnetically trapped bosonic $^{15}$NH($X\,^3\Sigma^-$) molecules, 
but the theory should
be general for any (ultra)cold paramagnetic species. We assume that the molecules are
in their vibrational and rotational ground states, as is the case experimentally \cite{campbell:07}.
For NH + NH, there 
are three spin-changing mechanisms that can induce trap loss: the
intramolecular spin-spin and spin-rotation couplings, and the intermolecular
magnetic dipolar coupling term \cite{krems:04b}. Previous theoretical work \cite{janssen:11a}
has shown that the intermolecular magnetic dipole interaction is the
main spin-relaxation mechanism for NH--NH at low collision energies and small to moderate
magnetic field strengths.
It was also shown, in the same paper,
that the dipolar spin-spin coupling term induces certain avoided crossings
between different adiabatic potential curves, which in turn give rise to spin-changing
transitions. That is, the spin-flip due to the intermolecular magnetic dipolar interaction
can be qualitatively understood in terms of the avoided curve crossings \cite{janssen:11a}.
In the present work, we discuss the influence of these crossings on the cross
section in much greater detail. We also provide analytical expressions for the
dipolar spin-relaxation cross section based on the Born approximation (BA) and distorted-wave
Born approximation (DWBA). We compare the analytical results with the cross sections
obtained from rigorous close-coupling (CC) calculations, and show that the results are in
excellent agreement over a wide range of collision energies and magnetic field strengths.

This paper is organized as follows. In Sec.\ \ref{subsec:qscatt}, we briefly describe the 
details of the CC calculations. The 
derivations of the BA and DWBA cross sections
are given in Secs.\ \ref{sec:BA} and \ref{sec:DWBA}, respectively, and the
results are discussed in Sec.\ \ref{sec:results}. The numerical results are
presented in Sec.\ \ref{sec:results_A}, with a particular emphasis on the role
of the avoided curve crossings, and the validity of the analytical BA and DWBA
cross sections is detailed in Sec.\ \ref{sec:results_B}. Finally, concluding
remarks are given in Sec.\ \ref{sec:concl}.

\section{Theory}
\label{sec:theory}

Throughout this paper, we will focus on collisions between two bosonic
$^{15}$NH($X\,^3\Sigma^-$) molecules in their magnetically trappable, low-field
seeking states $|S_A=1, M_{S_A}=1\rangle |S_B=1, M_{S_B}=1\rangle$.  Here
$S_{i}$ denotes the total electronic spin of the monomers ($i=A,B$) and
$M_{S_i}$ is the spin projection onto the magnetic field axis. A collision
complex of two such molecules is in the high-spin quintet $|S=2, M_S=2\rangle$
state, with $S$ denoting the total spin and $M_S$ its space-fixed projection.
Collisions that change either the $M_S$ quantum number of the quintet
state or the total spin $S$ to yield singlet ($S=0$) or triplet ($S=1$)
complexes will lead to immediate trap loss.

\subsection{Coupled-channel calculations}
\label{subsec:qscatt}
In order to obtain numerical values for the collision cross sections of
NH + NH, we have performed full CC calculations
as a function of energy and magnetic field. 
The details of these calculations are given elsewhere \cite{janssen:11a} and we provide
only a brief description here.  
The NH--NH scattering Hamiltonian is written as
\begin{equation}
\label{eq:H}
\hat{H} = -\frac{\hbar^2}{2\mu R} \frac{\partial^2}{\partial R^2}R +
          \frac{\hat{L}^2}{2\mu R^2} +
          V(\bm{R},\omega_A,\omega_B) +
          V_{\rm{magn.dip}}(\bm{R},\hat{\bm{S}}_A,\hat{\bm{S}}_B) +
          \hat{H}_A + \hat{H}_B,
\end{equation}
where $\mu$ is the reduced mass of the complex, 
$\bm{R}$ is the intermolecular vector that connects the centers of mass of the monomers,
$R=|\bm{R}|$,
$\hat{L}^2$ is the angular momentum operator associated with
rotation of $\bm{R}$, $V(\bm{R},\omega_A,\omega_B)$ is the potential-energy
surface for the quintet ($S=2$) state of NH--NH, $\omega_A$ and $\omega_B$ describe the orientation of
monomers $A$ and $B$,
$V_{\rm{magn.dip}}(\bm{R},\hat{\bm{S}}_A,\hat{\bm{S}}_B)$ is the intermolecular magnetic dipolar
interaction between the two spins, and $\hat{H}_A$ and $\hat{H}_B$ are
the Hamiltonians of the individual monomers. 
The magnetic dipole-dipole term is given
by
\begin{equation}
\label{eq:Vmagndip}
V_{\rm{magn.dip}}(\bm{R},\hat{\bm{S}}_A,\hat{\bm{S}}_B) = - \sqrt{6} g_S^2 \mu_{\rm{B}}^2 \frac{\alpha^2}{R^3}
                  \sum_q (-1)^q C_{2,-q}(\Omega_R) [\hat{\bm{S}}_A \otimes \hat{\bm{S}}_B]_q^{(2)},
\end{equation}
where $g_S \approx 2.0023$ is the electron $g$-factor, $\mu_{\rm{B}}$ is the Bohr magneton,
$\alpha$ is the fine-structure constant, 
$C_{2,-q}$ is a Racah-normalized spherical harmonic, 
$\Omega_R = (\Theta_R,\Phi_R)$ describes the orientation of $\bm{R}$,
and the factor in square brackets is the tensor product of the monomer spin operators
$\hat{\bm{S}}_A$ and $\hat{\bm{S}}_B$. 
The monomer operators $\hat{H}_i$ correspond to the asymptotic molecular states and 
account for the monomer rotation, intramolecular spin-spin coupling, spin-rotation coupling, and Zeeman interaction.
Hyperfine coupling is neglected.

The scattering calculations were carried out in a symmetry-adapted basis set that accounts for
the identical-particle symmetry of the system,
\begin{equation}
\label{eq:basis_sa}
| \phi^{\eta,\epsilon}_{\gamma_A \gamma_B L M_L} \rangle =
\frac{1}{[2(1+\delta_{\gamma_A \gamma_B})]^{1/2}}
\big[
| \gamma_A \gamma_B \rangle 
+ \eta (-1)^L
| \gamma_B \gamma_A \rangle \big] | L M_L \rangle.
\end{equation}
Here $\eta$ defines the symmetry of the wave function with respect to molecular interchange,
which is +1 for the bosonic $^{15}$NH -- $^{15}$NH complex, $\epsilon=(-1)^{N_A+N_B+L}$ is
the parity symmetry, which must be +1 for identical bosons in the same quantum state, 
and $|\gamma_A,\gamma_B\rangle$ 
denotes the molecular rotation and spin functions in the space-fixed frame \cite{krems:04b},
\begin{equation}
\label{eq:basis}
| \gamma_A \gamma_B \rangle
\equiv
| N_A M_{N_A} \rangle | S_A M_{S_A} \rangle
| N_B M_{N_B} \rangle | S_B M_{S_B} \rangle.
\end{equation}
The basis set was truncated at $N_A=N_B=2$ and $L=6$.
Although this basis set is not fully converged, we have verified that the calculated cross sections
are very similar to those obtained with $N_A=N_B=3$ in the region where the intermolecular
dipole-dipole coupling is dominant, i.e.\ at ultralow energies and small to moderate field strengths.
Increasing the rotational basis set does yield a larger cross-section contribution from the
\textit{intramolecular} spin-spin coupling, but this term becomes important only at energies above
$\sim$1 mK and fields above $\sim$100 G. For a more general discussion on the issue of basis-set
convergence, the reader is referred to Refs.\ \cite{janssen:11} and \cite{janssen:11a}.

Let us now consider the identical-particle symmetry of the complex. 
Even though hyperfine coupling is neglected, the symmetry of the nuclear-spin wave function should be
taken into account when evaluating the exchange symmetry of the total wave function.
We have assumed that both monomers are in their nuclear-spin stretched
states ($I=M_I=1$), so that the nuclear-spin function is symmetric under exchange.
Thus, we have $\eta=+1$ and $\epsilon=+1$.
We also point out that, due to parity conservation, collisions between rotational ground-state molecules
can only occur for \textit{even} values of $L$. 
Furthermore, the conservation of the
total angular momentum projection $\mathcal{M} = M_{N_A}+M_{N_B}+M_{S_A}+M_{S_B}+M_L$ requires
that any change in $M_{S_A}$ or $M_{S_B}$ must be accompanied by a change in $M_L$.
It therefore follows that, in the ultracold regime, the $s$-wave spin-inelastic collision channel
for magnetically trapped, rotational ground-state NH is dominated by the $L=2$ outgoing partial wave.

We performed the scattering calculations for each value of $\mathcal{M}$
and accumulated the resulting scattering $S$-matrices to extract the cross sections. 
The calculations were carried out using 
a modified version of the MOLSCAT package \cite{molscat:94,gonzalez:07}.
The propagation was performed using the hybrid log-derivative method of Alexander and Manolopoulos \cite{alexander:87}.
Prior to matching to asymptotic boundary conditions,
an additional transformation was required to obtain the exact channel eigenfunctions \cite{krems:04b}. This is because
the intramolecular spin-spin coupling mixes states with $N_i$ and $N_i\pm2$, which makes $N_i$, $M_{N_i}$, and $M_{S_i}$
only approximately good quantum numbers.
The \textit{exact} molecular eigenstates will be denoted as
\begin{equation}
\label{eq:channelbasis}
|\bar{\gamma}_A \bar{\gamma}_B\rangle
\equiv
| (\bar{N}_A,S_A) J_A, M_{J_A} \rangle
| (\bar{N}_B,S_B) J_B, M_{J_B} \rangle.
\end{equation}
We emphasize that the intramolecular coupling is relatively weak and $N_i$, $M_{N_i}$, and $M_{S_i}$
may be treated as almost exact. Specifically, for the rotational ground state of $^{15}$NH,
the magnetically trapped component with $J_i=M_{J_i}=1$
contains 99.992\% of $|N_i=0, M_{N_i}=0, S_i=1, M_{S_i}=1\rangle$.

\subsection{Born approximation}
\label{sec:BA}
In this section, we derive an analytical expression for the inelastic
spin-changing cross section due to $V_{\rm{magn.dip}}$ based on the first-order Born
approximation. This approximation assumes that the interaction between
projectile and target is so weak that the initial and final states can be
described by undistorted plane waves. We note that the BA has been previously used
in the study of cold collisions in e.g.\ Refs.\ \cite{moerdijk:96,avdeenkov:05,kajita:06,zygelman:10}.
The aim of the
present work is to give a cross-section expression in closed form,
and we therefore outline the complete derivation for the sake of clarity.
The derived expression is general for any two paramagnetic species for which the
electronic spin and its space-fixed projection are (approximately) good quantum numbers,
e.g.\ for Hund's case (b) molecules and $S$-state atoms,
but we will apply it only to the case of NH($^3\Sigma^-$) + NH($^3\Sigma^-$).

We start with the exact expression for the differential cross section
(see e.g.\ Eq.\ (XIX.19) of Ref.\ \cite{messiah:69}),
\begin{equation}
\label{eq:diffxsec}
\frac{d\sigma_{a \rightarrow b} (\Omega_a)}{d\Omega_b} = 
\frac{2\pi}{\hbar v_a}
\big| \big\langle \bm{k}_b, \bar{\gamma}^{(b)}_{A}, \bar{\gamma}^{(b)}_{B}
\big| V_{\rm{int}}
\big| \bm{k}_a^{(+)},\bar{\gamma}^{(a)}_{A},\bar{\gamma}^{(a)}_{B} \big\rangle
\big|^2 \rho_b(E),
\end{equation}
where $a$ and $b$ label the initial and final states, respectively,
$\Omega_a = (\theta_{k_a},\phi_{k_a})$ and $\Omega_b = (\theta_{k_b},\phi_{k_b})$
describe the directions of the incoming and outgoing collision fluxes,
$\bm{k}_a^{(+)}$ is the exact incident wave function with wavenumber $k_a$,
$\bm{k}_b$ is a plane wave with wavenumber $k_b$,
$\bar{\gamma}^{(i)}_{A}$ and $\bar{\gamma}^{(i)}_{B}$
denote the internal quantum numbers of the monomers for the initial and final states ($i=a,b$),
$V_{\rm{int}}$ is the interaction between the
scattering particles, for which we take $V_{\rm{int}} = V_{\rm{magn.dip}}$,
$v_a=\hbar k_a/\mu$ is the velocity of the incident beam,
and $\rho_b(E) = \mu k_b /[\hbar^2 (2\pi)^{3}]$ is the density of final states at energy $E=\frac{1}{2} \mu v_a^2$.
The first-order Born approximation amounts to approximating the incident wave function 
as a plane wave, i.e.\ $|\bm{k}_a^{(+)} \rangle \approx |\bm{k}_a \rangle$.
The plane waves are normalized to unit density and are mutually orthogonal,
\begin{eqnarray}
|\bm{k}\rangle &=& e^{i\bm{k}\cdot\bm{R}}, \\
\langle \bm{k} | \bm{k}' \rangle &=& (2\pi)^3 \delta(\bm{k}-\bm{k}').
\end{eqnarray}
Here $\delta(\bm{k}-\bm{k}')$ represents the three-dimensional Dirac delta function.

In the case of NH + NH, the asymptotic states $\bar{\gamma}^{(i)}_{A}$ and
$\bar{\gamma}^{(i)}_{B}$ should be described as in Eq.\ (\ref{eq:channelbasis}).
However, since we focus on collisions between rotational ground-state molecules,
we may treat $M_{S_A}$ and $M_{S_B}$ as almost exact quantum numbers. Furthermore, 
taking into account that $V_{\rm{magn.dip}}$ acts only on the vector $\bm{R}$ and the
electron-spin coordinates, we can omit the molecular rotational quantum numbers
and write
\begin{eqnarray}
\label{eq:spinfcions}
\big| \bar{\gamma}_A^{(i)} \big\rangle \approx \big| S_A M_{S_A}^{(i)} \big\rangle, \nonumber \\
\big| \bar{\gamma}_B^{(i)} \big\rangle \approx \big| S_B M_{S_B}^{(i)} \big\rangle.
\end{eqnarray}
The energies of the initial and final molecular states are now determined only by their 
Zeeman shifts. 
If we define the Zeeman levels relative to the initial state,
the wavenumbers are
$k_a = \sqrt{2\mu E}/\hbar$ and
$k_b = \sqrt{2\mu(E + g_S\mu_{\rm{B}} \Delta M_S B)}/\hbar$,
where $\Delta M_S$ is the total spin-change,
$\Delta M_S = M_{S_A}^{(a)} + M_{S_B}^{(a)} - M_{S_A}^{(b)} - M_{S_B}^{(b)}$,
the term $g_S\mu_{\rm{B}} \Delta M_S B$ is the corresponding change in Zeeman energy, and $B$ is the magnetic field strength.
In the remainder of this paper,
we will use $B$ exclusively to indicate the magnetic field strength, while the
subscript $B$ is used to label the quantum numbers of monomer $B$.

The plane waves can be expanded in terms of partial waves as 
\begin{equation}
\label{eq:pwexpansion}
e^{i\bm{k}\cdot\bm{R}} = 4\pi \sum_{L=0}^{\infty} \sum_{M=-L}^{L}
i^L j_L(kR) Y_{L,M}(\Theta_R,\Phi_R) Y^{*}_{L,M}(\theta_k,\phi_k),
\end{equation}
where $j_L(kR)$ is a spherical Bessel function of the first kind, the $Y_{L,M}$ functions are
spherical harmonics, and the superscript * denotes complex conjugation.
If we now substitute Eq.\ (\ref{eq:Vmagndip}) for the particle interaction 
and use Eq.\ (\ref{eq:spinfcions})
to describe the molecular asymptotic states, we obtain
\begin{eqnarray}
\label{eq:matelement}
\lefteqn{
\big\langle \bm{k}_b, \bar{\gamma}^{(b)}_{A}, \bar{\gamma}^{(b)}_{B}
\big| V_{\rm{magn.dip}}
\big| \bm{k}_a,\bar{\gamma}^{(a)}_{A},\bar{\gamma}^{(a)}_{B} \big\rangle = 
    -4\pi \sqrt{6} g_S^2 \mu_{\rm{B}}^2 \alpha^2 
    \sum_{L_a} \sum_{L_b} i^{L_a-L_b} } \nonumber \\
& &  \times \sum_{M_b} Y_{L_b,M_b}(\Omega_b) 
            \sum_{M_a} Y^{*}_{L_a,M_a}(\Omega_a) 
\sum_q (-1)^q 
    \int_R 
     j_{L_b}(k_bR) \frac{1}{R^3} j_{L_a}(k_aR) R^2 dR \nonumber \\
& & \times 
    \int_{\Omega_R} 
    Y_{L_b,M_b}^*(\Omega_R) C_{2,-q}(\Omega_R) Y_{L_a,M_a}(\Omega_R)
    d\Omega_R \hphantom{x} \nonumber \\
& & \times 
     \big\langle S_A M_{S_A}^{(b)}, S_B M_{S_B}^{(b)} 
     \big|  [\hat{\bm{S}}_A \otimes \hat{\bm{S}}_B]_q^{(2)}
     \big| S_A M_{S_A}^{(a)}, S_B M_{S_B}^{(a)} \big\rangle,
\end{eqnarray}
where the last factor represents an integral over the spin coordinates.
The integral over $R$ can be performed analytically and gives, for $k_a \leq k_b$
(see also Ref.\ \cite{avdeenkov:05}),
\begin{eqnarray}
\label{eq:intR_BA}
\lefteqn{ \int j_{L_b}(k_bR) \frac{1}{R} j_{L_a}(k_aR) dR = 
  \frac{\pi}{8} \left(\frac{k_a}{k_b}\right)^{L_a} } \nonumber \\
& & \hphantom{xxx}
    \times \frac{\Gamma\left(\frac{L_a+L_b}{2}\right)}
                {\Gamma\left(\frac{L_b-L_a+3}{2}\right) \Gamma\left(L_a+\frac{3}{2}\right)}  \nonumber \\
& & \hphantom{xxx}
    \times {}_2F_1\left(\frac{L_a-L_b-1}{2}, \frac{L_a+L_b}{2}, L_a + \frac{3}{2},\frac{k_a^2}{k_b^2}\right),
\end{eqnarray}
where $\Gamma$ is the Gamma function and ${}_2F_1$ is Gauss's hypergeometric function.
The integral over $\Omega_R$ gives
\begin{eqnarray}
\lefteqn{
  \int_{\Omega_R} Y_{L_b,M_b}^*(\Omega_R) C_{2,-q}(\Omega_R) Y_{L_a,M_a}(\Omega_R)
  d\Omega_R =} \nonumber \\
& & \hphantom{xx}
    \sqrt{(2L_a+1)(2L_b+1)} (-1)^{M_b}
    \left( \begin{array}{ccc} L_b & 2 & L_a \\ 0 & 0 & 0 \end{array} \right)
    \left( \begin{array}{ccc} L_b & 2 & L_a \\ -M_b & -q & M_a \end{array} \right),
    \hphantom{xxx}
\end{eqnarray}
with the terms in large round brackets denoting Wigner 3$j$ symbols. 
The last 3$j$ symbol readily implies that $q=M_a-M_b$. Finally, for the spin-dependent
term we find
\begin{eqnarray}
\label{eq:matspin}
\lefteqn{
  \big\langle S_A M_{S_A}^{(b)}, S_B M_{S_B}^{(b)}
  \big|  [\hat{\bm{S}}_A \otimes \hat{\bm{S}}_B]_q^{(2)}
   \big| S_A M_{S_A}^{(a)}, S_B M_{S_B}^{(a)} \big\rangle = } \nonumber \\
& & \sqrt{5} (-1)^{q+S_A+S_B-M_{S_A}^{(b)}-M_{S_B}^{(b)}} \nonumber \\
& & \times [S_A(S_A+1)(2S_A+1) S_B(S_B+1)(2S_B+1)]^{1/2} \nonumber \\
& & \times \sum_{M_A',M_B'} 
    \left( \begin{array}{ccc} 1 & 1 & 2 \\ M_A' & M_B' & -q \end{array} \right)
    \left( \begin{array}{ccc} S_A & 1 & S_A \\ -M_{S_A}^{(b)} & M_A' & M_{S_A}^{(a)} \end{array} \right)
    \left( \begin{array}{ccc} S_B & 1 & S_B \\ -M_{S_B}^{(b)} & M_B' & M_{S_B}^{(a)} \end{array} \right).
\end{eqnarray}
Note that the sums over $M_A'$ and $M_B'$ collapse for given values of
$M_{S_A}^{(i)}$ and $M_{S_B}^{(i)}$, since the last two 3$j$ symbols require
that $M_A' = M_{S_A}^{(b)} - M_{S_A}^{(a)}$ and $M_B' = M_{S_B}^{(b)} -
M_{S_B}^{(a)}$. Furthermore, we have $M_A'+M_B'=q$ so that
$M_b-M_a = M_{S_A}^{(a)}+M_{S_B}^{(a)}-M_{S_A}^{(b)}- M_{S_B}^{(b)} = \Delta M_S$.
The sums over $M_a$, $M_b$ and $q$ in Eq.\ (\ref{eq:matelement}) therefore reduce
to a single sum 
for any individual matrix element. 
The differential cross section is now
readily calculated by substituting Eqs.\ (\ref{eq:matelement}) -- (\ref{eq:matspin}) into Eq.\
(\ref{eq:diffxsec}).

The integral cross section is obtained by integrating 
$d\sigma_{a \rightarrow b}/d\Omega_b$ over all orientations of the outgoing wave
and averaging over all directions of the incoming collision flux,
\begin{equation}
\label{eq:intxsec}
\sigma_{a\rightarrow b}(E) = \frac{1}{4\pi} \int_{\Omega_a} d\Omega_a
                             \int_{\Omega_b}
\frac{d\sigma_{a \rightarrow b}(\Omega_a) }{d\Omega_b} d\Omega_b.
\end{equation}
Using the orthogonality relation
$\int Y^*_{L,M}(\Omega) Y_{L',M'}(\Omega) d\Omega = \delta_{L,L'} \delta_{M,M'}$, we find the
following expression for the BA cross section 
for $|S_A M_{S_A}^{(a)}, S_B M_{S_B}^{(a)} \rangle
\rightarrow |S_A M_{S_A}^{(b)}, S_B M_{S_B}^{(b)}\rangle$ transitions induced by $V_{\rm{magn.dip}}$: 
\begin{eqnarray}
\label{eq:sigBA}
\lefteqn{
\sigma^{\rm{BA}}_{a\rightarrow b}(E) = 
 \frac{15\pi^3}{2\hbar^4} 
 \mu^2 g_S^4\mu_{\rm{B}}^4 \alpha^4
 \sum_{L_a} \sum_{L_b}
 (2L_a+1) (2L_b+1) 
 \left(\frac{k_a}{k_b}\right)^{2L_a-1} } \nonumber \\
& & \times S_A(S_A+1)(2S_A+1) S_B(S_B+1)(2S_B+1) \nonumber \\
& & \times \left[ 
               \frac{\Gamma\left(\frac{L_a+L_b}{2}\right)}
               {\Gamma\left(\frac{L_b-L_a+3}{2}\right) \Gamma\left(L_a+\frac{3}{2}\right)} \right]^2  \nonumber \\
& & \times\left[ {}_2F_1\left(\frac{L_a-L_b-1}{2}, \frac{L_a+L_b}{2}, L_a + \frac{3}{2},\frac{k_a^2}{k_b^2}\right) \right]^2 \nonumber \\
& & \times 
      \sum_{M_a}
      \left( \begin{array}{ccc} L_b & 2 & L_a \\ -(M_a+\Delta M_S) & \Delta M_S & M_a \end{array} \right)^2 \nonumber \\
& & \times \left[
      \left( \begin{array}{ccc} L_b & 2 & L_a \\ 0 & 0 & 0 \end{array} \right)
      \left( \begin{array}{ccc} 1 & 1 & 2 \\ \Delta M_{S_A} & \Delta M_{S_B} & \Delta M_S \end{array} \right)
      \right]^2 \nonumber \\
& & \times\left[ 
      \left( \begin{array}{ccc} S_A & 1 & S_A \\ -M_{S_A}^{(b)} & \Delta M_{S_A} & M_{S_A}^{(a)} \end{array} \right)
      \left( \begin{array}{ccc} S_B & 1 & S_B \\ -M_{S_B}^{(b)} & \Delta M_{S_B} & M_{S_B}^{(a)} \end{array} \right) 
      \right]^2,
\end{eqnarray}
with $\Delta M_{S_A} = M_{S_A}^{(b)} - M_{S_A}^{(a)}$ and $\Delta M_{S_B} = M_{S_B}^{(b)} - M_{S_B}^{(a)}$.
The cross section for a specific incoming partial wave $L_a$ and a certain outgoing wave $L_b$ is
obtained by simply omitting the sums over $L_a$ and $L_b$.
We also point out that in the limit of $k_a \ll k_b$,
which holds for ultracold exothermic collisions,
the hypergeometric function ${}_2F_1$ becomes 1 (see e.g.\ Eq.\ 15.1.1 of Ref.\ \cite{abramowitz:64})
and the energy dependence
of the cross section is $\sigma_{a\rightarrow b} \sim k_a^{2L_a-1} \sim
E^{L_a-1/2}$. The cross-section behaviour as a function of $B$ is then, for $B \gg E/(g_S \mu_{\rm{B}}\Delta M_S)$,
$\sigma_{a\rightarrow b} \sim k_b^{1-2L_a} \sim B^{1/2 - L_a}$. 
Note that this $B$-dependence is different from the threshold law derived by Volpi and Bohn \cite{volpi:02}.
They considered the case of spin-changing transitions induced \textit{inside} the centrifugal barrier of the exit
channel, and found that the cross section behaves as $\sigma_{a\rightarrow b} \sim B^{L_b+1/2}$.
In our case, however, the spin-flip takes place at long range, \textit{outside} the centrifugal barrier,
and hence we find a different result. The long-range mechanism for dipolar spin relaxation will be addressed
in detail in Sec.\ \ref{sec:results_A}. 

Equation (\ref{eq:sigBA}) is valid for any paramagnetic species that can be represented as in Eq.\ (\ref{eq:spinfcions}).
We note that, in the case of identical particles, 
the cross section must be multiplied by a 
factor of 2 if both monomers are in the same initial state, i.e. if $M_{S_A}^{(a)}=M_{S_B}^{(a)}$
(see e.g.\ Appendix B of Ref.\ \cite{tscherbul:09c}). This also applies to collisions between two
magnetically trapped NH molecules, for which $M_{S_A}^{(a)}=M_{S_B}^{(a)}=1$.

\subsection{Analytical distorted-wave Born approximation}
\label{sec:DWBA}
As will be shown in Sec.\ \ref{sec:results}, the first-order BA is very accurate
at low collision energies, but starts to deviate from the CC result at high energies
and strong magnetic fields. One of the causes for this discrepancy is the phase shift
in the incoming scattering channel. In order to quantify this effect, we have developed
an analytical distorted-wave Born approximation in which the phase shift in the incident
plane wave is explicitly included.

Our starting point for the analytical DWBA is again Eq.\ (\ref{eq:diffxsec}),
but now we approximate the incoming wave function $| \bm{k}_a^{(+)} \rangle$
as an elastically distorted wave $| \bm{k}_a' \rangle$,
\begin{equation}
\label{eq:ka_DWBA}
| \bm{k}_a' \rangle = 4\pi \sum_{L_a} \sum_{M_a} i^{L_a} \frac{1}{2}\big[h^{(2)}_{L_a}(k_aR) + S_{aa}^{(M_a)}h^{(1)}_{L_a}(k_aR)\big] Y_{L_a,M_a}(\Theta_R,\Phi_R) Y^{*}_{L_a,M_a}(\theta_{k_a},\phi_{k_a}),
\end{equation}
where $h^{(1)}_{L_a}$ and $h^{(2)}_{L_a}$ are spherical Hankel functions of the
first and second kind, respectively, and $S_{aa}^{(M_a)}$ is the elastic
$S$-matrix element that contains the phase shift for the incident scattering
channel $|S_A M_{S_A}^{(a)}, S_B M_{S_B}^{(a)}, L_a, M_a \rangle$.
The $S^{(M_a)}_{aa}$ matrix elements for NH--NH are taken from the full CC calculations 
described in Sec.\ \ref{subsec:qscatt}.
The Hankel functions are defined in terms of regular and irregular spherical Bessel functions as
\begin{eqnarray}
\label{eq:defHankel}
h^{(1)}_{L}(z) &=& j_L(z) + iy_L(z), \nonumber \\
h^{(2)}_{L}(z) &=& j_L(z) - iy_L(z),
\end{eqnarray}
where $y_L(z)$ is a spherical Bessel function of the second kind.
We note that the wave function of Eq.\ (\ref{eq:ka_DWBA}) is unphysically divergent at the origin
for nonzero phase shifts ($S_{aa}^{(M_a)} \neq 1$), 
but matches the (exact) CC wave function at sufficiently large $R$.
Hence, the approximation of
$| \bm{k}_a^{(+)} \rangle \approx | \bm{k}_a' \rangle$ constitutes an improvement over
the first-order BA if the coupling occurs at long range. 

The radial part of Eq.\ (\ref{eq:ka_DWBA})
may also be written in terms of the transmission matrix element $T^{(M_a)}_{aa} = 1 - S^{(M_a)}_{aa}$,
\begin{eqnarray}
\label{eq:Tmat}
\frac{1}{2}\big[ h^{(2)}_{L_a}(k_aR) + S_{aa}h^{(1)}_{L_a}(k_aR) \big]
   &=& \frac{1}{2}\big[ h^{(2)}_{L_a}(k_aR) + h^{(1)}_{L_a}(k_aR) - T^{(M_a)}_{aa}h^{(1)}_{L_a}(k_aR) \big] \nonumber \\
   &=& j_{L_a}(k_aR) - \frac{1}{2}T^{(M_a)}_{aa}h^{(1)}_{L_a}(k_aR). 
\end{eqnarray}
Substitution of Eqs.\ (\ref{eq:ka_DWBA}) and (\ref{eq:Tmat}) into (\ref{eq:matelement}) for the matrix
element over $V_{\rm{magn.dip}}$ gives a radial integral of the form
\begin{equation}
\label{eq:intR_DWBA1}
\int j_{L_b}(k_bR) \frac{1}{R} j_{L_a}(k_aR) dR - \frac{1}{2}T^{(M_a)}_{aa} \int j_{L_b}(k_bR) \frac{1}{R} h_{L_a}^{(1)}(k_aR) dR.
\end{equation}
Note that the first integral is identical to that of Eq.\ (\ref{eq:intR_BA}). Using Eq.\ (\ref{eq:defHankel}),
we may write the second integral of Eq.\ (\ref{eq:intR_DWBA1}) as
\begin{eqnarray}
\label{eq:intR_DWBA2}
\lefteqn{ \int j_{L_b}(k_bR) \frac{1}{R} h_{L_a}^{(1)}(k_aR) dR = } \nonumber \\
& & \hphantom{xxx} \int j_{L_b}(k_bR) \frac{1}{R} j_{L_a}(k_aR) dR + i \int j_{L_b}(k_bR) \frac{1}{R} y_{L_a}(k_aR) dR.
\end{eqnarray}
Again we observe that the first integral on the right-hand side is given by Eq.\ (\ref{eq:intR_BA}).
The second integral on the right-hand side is convergent only for $L_b > L_a + 1$ and gives,
for $k_a < k_b$ and integer $L_a$ and $L_b$,
\begin{eqnarray}
\label{eq:intR_DWBA3}
\lefteqn{
  \int j_{L_b}(k_bR) \frac{1}{R} y_{L_a}(k_aR) dR = \frac{-1}{8} \left(\frac{k_b}{k_a}\right)^{L_a+1}  } \nonumber \\
& & \times \frac{\Gamma\left(L_a+\frac{1}{2}\right) \Gamma\left(\frac{L_b-L_a-1}{2}\right) }
                {\Gamma\left(\frac{L_a+L_b+4}{2}\right) } \nonumber \\
& & \times {}_2F_1\left(\frac{-L_a-L_b-2}{2}, \frac{L_b-L_a-1}{2}, \frac{1}{2}-L_a, \frac{ k_a^2}{k_b^2}\right).
\end{eqnarray}
We can now replace the radial integral in Eq.\ (\ref{eq:matelement}) by the
expression of Eq.\ (\ref{eq:intR_DWBA1}) to obtain the matrix element over
$V_{\rm{magn.dip}}$ in our distorted-wave Born approximation. Substitution into
Eq.\ (\ref{eq:diffxsec}) gives the differential DWBA cross section, and Eq.\
(\ref{eq:intxsec}) subsequently yields the integral cross section. The final
expression for the DWBA spin-inelastic cross section due to $V_{\rm{magn.dip}}$ is
\begin{eqnarray}
\label{eq:sigDWBA}
\lefteqn{
\sigma^{\rm{DWBA}}_{a\rightarrow b}(E) = 
 \frac{15\pi}{2\hbar^4} 
 \mu^2 g_S^4\mu_{\rm{B}}^4 \alpha^4
 \sum_{L_a} \sum_{L_b}
 (2L_a+1) (2L_b+1) 
 \frac{k_b}{k_a} } \nonumber \\
& & \times S_A(S_A+1)(2S_A+1) S_B(S_B+1)(2S_B+1) \nonumber \\
& & \times \sum_{M_a}
           \left| \left(1 - \frac{1}{2}T^{(M_a)}_{aa}\right) \pi \left(\frac{k_a}{k_b}\right)^{L_a}
               \frac{\Gamma\left(\frac{L_a+L_b}{2}\right)}
               {\Gamma\left(\frac{L_b-L_a+3}{2}\right) \Gamma\left(L_a+\frac{3}{2}\right)} \right. \nonumber \\
& & \times {}_2F_1\left(\frac{L_a-L_b-1}{2}, \frac{L_a+L_b}{2}, L_a + \frac{3}{2},\frac{k_a^2}{k_b^2}\right) 
           + i\frac{1}{2}T_{aa}^{(M_a)} \left(\frac{k_b}{k_a}\right)^{L_a+1}  \nonumber \\
& & \times \left. \frac{\Gamma\left(L_a+\frac{1}{2}\right) \Gamma\left(\frac{L_b-L_a-1}{2}\right) }
                {\Gamma\left(\frac{L_a+L_b+4}{2}\right) }
           {}_2F_1\left(\frac{-L_a-L_b-2}{2}, \frac{L_b-L_a-1}{2}, \frac{1}{2}-L_a, \frac{ k_a^2}{k_b^2}\right) \right|^2 \nonumber \\
& & \times
      \left( \begin{array}{ccc} L_b & 2 & L_a \\ -(M_a+\Delta M_S) & \Delta M_S & M_a \end{array} \right)^2 \nonumber \\
& & \times\left[
      \left( \begin{array}{ccc} L_b & 2 & L_a \\ 0 & 0 & 0 \end{array} \right)
      \left( \begin{array}{ccc} 1 & 1 & 2 \\ \Delta M_{S_A} & \Delta M_{S_B} & \Delta M_S \end{array} \right)
      \right]^2 \nonumber \\
& & \times\left[ 
      \left( \begin{array}{ccc} S_A & 1 & S_A \\ -M_{S_A}^{(b)} & \Delta M_{S_A} & M_{S_A}^{(a)} \end{array} \right)
      \left( \begin{array}{ccc} S_B & 1 & S_B \\ -M_{S_B}^{(b)} & \Delta M_{S_B} & M_{S_B}^{(a)} \end{array} \right)
      \right]^2.
\end{eqnarray}
The BA result of Eq.\ (\ref{eq:sigBA}) is recovered in the limit of
$T^{(M_a)}_{aa}\rightarrow 0$. We emphasize that, in contrast to the BA, the sums over $L_a$ and $L_b$
in our DWBA expression should be restricted such that $L_b>L_a+1$ [see Eq.\ (\ref{eq:intR_DWBA3})].
We also note again that, for indistinguishable
particles such as NH + NH, the cross section must be multiplied by 2 if the
monomers are in the same initial state.

\section{Results and discussion}
\label{sec:results}

\subsection{Numerical results}
\label{sec:results_A}
We first discuss the numerical results for NH--NH obtained from full CC calculations. 
Previous theoretical work \cite{janssen:11a} has shown that the intermolecular
magnetic dipole interaction is the dominant trap-loss mechanism for NH--NH
at low collision energies and small magnetic fields, while at higher energies and fields
the intramolecular couplings become increasingly important. Here we will address only
the intermolecular coupling term and provide a careful analysis of its contribution to
the total inelastic cross section.

As explained in Ref.\ \cite{janssen:11a}, the contribution from
$V_{\rm{magn.dip}}$ is most easily understood by considering the adiabatic
potential curves. We will repeat part of this discussion here for the sake of clarity. 
Asymptotically, the adiabatic curves correspond to the molecular
eigenstates $\bar{\gamma}_A$ and $\bar{\gamma}_B$, and at finite $R$ each curve
also contains a centrifugal barrier (see Fig.\ 1 of Ref.\ \cite{janssen:11a}).
Thus, at long range, the adiabats can be labeled by
$|\bar{\gamma}_A,\bar{\gamma}_B\rangle$ and $L$, and therefore also correlate
to scattering channels.  It has already been noted in Refs.\ \cite{wallis:10}
and \cite{janssen:11a} that several adiabatic curves are narrowly avoided due
to the intermolecular magnetic dipole interaction, and that the spin-flip
induced by $V_{\rm{magn.dip}}$ takes place at the corresponding crossing.  If
we neglect the weak intramolecular spin-spin and spin-rotation couplings so
that Eq.\ (\ref{eq:spinfcions}) holds, we can define the
avoided-crossing points $R_c$ as
\begin{equation}
\label{eq:Rc}
g_S \mu_{\rm{B}} B \Delta M_S = \frac{\hbar^2 \left[L_b(L_b+1) - L_a(L_a+1)\right] }{2\mu R_c^2},
\end{equation}
where $L_a$ and $L_b$ denote the values of $L$ for the adiabats correlating to
the incoming and outgoing channels, respectively.  The energies at which the
crossings occur are given by
\begin{equation}
\label{eq:Ec}
E_c = \frac{\hbar^2 L_a(L_a+1)}{2\mu R_c^2},
\end{equation}
defined relative to the threshold of the incident channel.  We must point out
that, since $V_{\rm{magn.dip}}$ contains a second-rank tensor in $\Omega_R$ and
first-rank tensors in the monomer spin coordinates, the avoided crossings occur
only if $L_a$ and $L_b$ differ at most by 2 and $M_{S_i}^{(a)}$ and
$M_{S_i}^{(b)}$ ($i=A,B$) each differ at most by 1.  Thus, not all crossings
are avoided.

It can be deduced from Eq.\ (\ref{eq:Rc}) that, for small to moderate field
strengths, the crossing points $R_c$ are located at very long range. Therefore,
the spin-change due to $V_{\rm{magn.dip}}$ can occur without having to overcome
the centrifugal barrier in the outgoing channel.  More specifically, the
$|M_{S_A} = M_{S_B}=1, L_a=0\rangle$ incident channel of NH--NH can couple with
the $|M_{S_A} =1, M_{S_B}=0, L_b=2\rangle$ and $|M_{S_A} = M_{S_B}=0,
L_b=2\rangle$ outgoing channels even at zero collision energy.  Hence, at low
energies and relatively low magnetic fields, the intermolecular magnetic
dipolar interaction is the main source of trap loss for NH--NH.

\begin{figure}
\centering
\includegraphics[width=15cm]{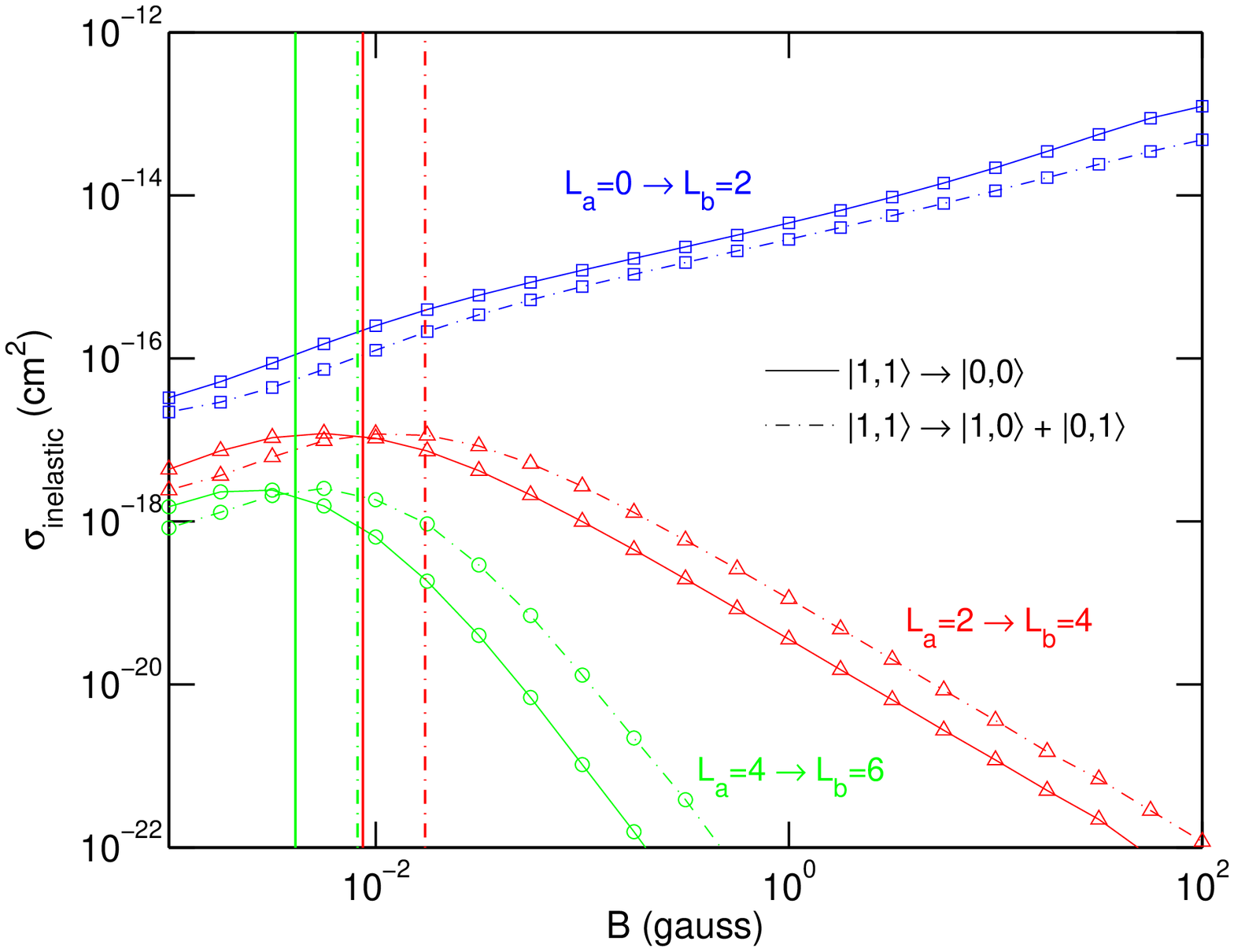}
\caption
  {\label{fig:pw_1muK}
State-to-state collision cross sections for the
$| M_{S_A} = 1, M_{S_B} = 1\rangle \rightarrow | M_{S_A} = 0, M_{S_B} = 0\rangle$ and
$| M_{S_A} = 1, M_{S_B} = 1\rangle \rightarrow | M_{S_A} = 1, M_{S_B} = 0\rangle + |M_{S_A} = 0, M_{S_B} = 1\rangle$
transitions of NH + NH,
obtained from full CC calculations as a function of magnetic field at $E = 10^{-6}$ K.
The (blue) lines marked with squares correspond to $L_a=0 \rightarrow L_b=2$ transitions, the (red) lines
marked with triangles correspond to $L_a=2 \rightarrow L_b=4$, and the (green) lines marked with circles
correspond to $L_a=4 \rightarrow L_b=6$.
The vertical lines indicate the $B_c$ values below which the crossings for
$L_a=2 \rightarrow L_b = 4$ and $L_a=4 \rightarrow L_b = 6$ are energetically accessible.
}
\end{figure}

\begin{figure}
\centering
\includegraphics[width=15cm]{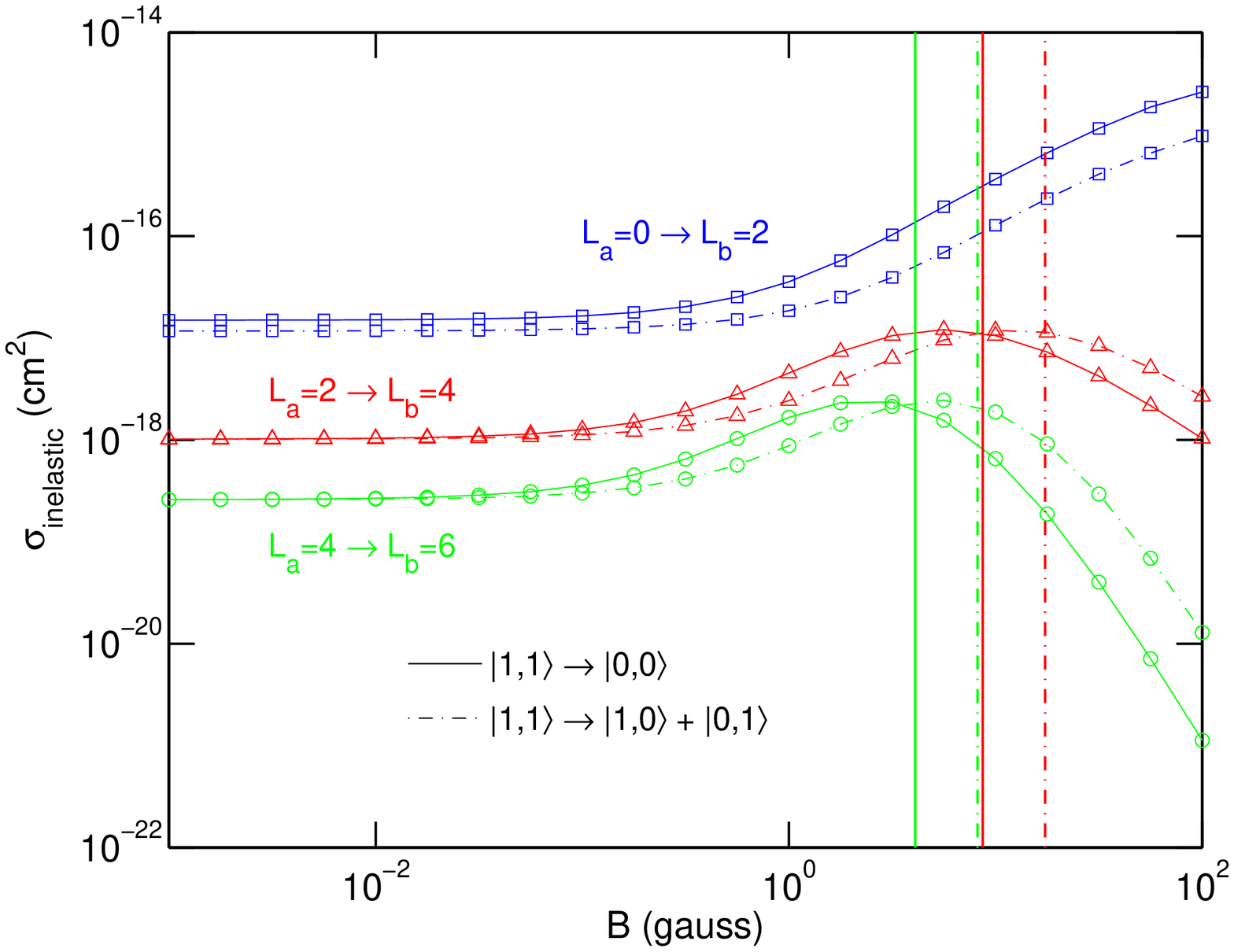}
\caption
  {\label{fig:pw_1mK}
State-to-state collision cross sections for the
$| M_{S_A} = 1, M_{S_B} = 1\rangle \rightarrow | M_{S_A} = 0, M_{S_B} = 0\rangle$ and
$| M_{S_A} = 1, M_{S_B} = 1\rangle \rightarrow | M_{S_A} = 1, M_{S_B} = 0\rangle + |M_{S_A} = 0, M_{S_B} = 1\rangle$
transitions of NH + NH,
obtained from full CC calculations as a function of magnetic field at $E=10^{-3}$ K.
The (blue) lines marked with squares correspond to $L_a=0 \rightarrow L_b=2$ transitions, the (red) lines
marked with triangles correspond to $L_a=2 \rightarrow L_b=4$, and the (green) lines marked with circles
correspond to $L_a=4 \rightarrow L_b=6$.
The vertical lines indicate the $B_c$ values below which the crossings for
$L_a=2 \rightarrow L_b = 4$ and $L_a=4 \rightarrow L_b = 6$ are energetically accessible.
}
\end{figure}

It also follows from Eqs.\ (\ref{eq:Rc}) and (\ref{eq:Ec}) that the curve crossings for higher
partial waves, e.g.\ for the $|M_{S_A}=M_{S_B}=1, L_a=2\rangle \rightarrow
|M_{S_A}=M_{S_B}=0, L_b=4\rangle$ transition, become \textit{lower} in energy
as the magnetic field \textit{decreases}. This implies that, at a fixed
collision energy, the $L_a=2 \rightarrow L_b=4$ channel transitions open up \textit{below}
a certain $B$-value. We will denote this critical magnetic field strength as
$B_c$. 
Figures \ref{fig:pw_1muK} and \ref{fig:pw_1mK} show the state-to-state inelastic NH--NH cross sections for different
$L_a \rightarrow L_b$ channel transitions as a function of $B$ at collision energies of 10$^{-6}$ K and 10$^{-3}$ K, respectively.
The $B_c$ values for the $L_a=2 \rightarrow L_b=4$ and 
$L_a=4 \rightarrow L_b=6$ crossings are also indicated. For the $|M_{S_A}=M_{S_B}=1\rangle \rightarrow |M_{S_A}=M_{S_B}=0\rangle$
transitions, with $\Delta M_S = 2$, the numerical values are $B_c=8.67\times10^{-3}$ G at 10$^{-6}$ K and
$B_c=4.09\times10^{-3}$ G at 10$^{-3}$ K for $L_a = 2 \rightarrow L_b = 4$, and
$B_c=8.67$ G at 10$^{-6}$ K and
$B_c=4.09$ G at 10$^{-3}$ K for $L_a = 4 \rightarrow L_b = 6$.
The critical field strengths for the $|M_{S_A}=M_{S_B}=1\rangle \rightarrow
|M_{S_A}=1,M_{S_B}=0\rangle + |M_{S_A}=0,M_{S_B}=1\rangle$ transitions, with $\Delta M_S = 1$, are twice as large as those for $\Delta M_S = 2$.
It can be seen that 
the inelastic cross sections for $L_a=2\rightarrow L_b=4$ and $L_a=4\rightarrow L_b=6$
indeed decrease as $B$ exceeds the corresponding $B_c$ value. 
This $B$-dependence is remarkable, 
considering that higher partial waves typically contribute only if the
exothermicity in the outgoing channel is large. Due to the long-range nature of
the magnetic dipole interaction, however, the scattering of higher partial
waves becomes increasingly important as the exothermicity \textit{decreases}, because
the crossings points $R_c$ occur at a larger distance.

The influence of the kinetic energy on the inelastic cross section can also be understood
in terms of the adiabatic curve crossings.
For a given magnetic field strength, the avoided
crossings for $L_a =2 \rightarrow L_b = 4$ and $L_a =4 \rightarrow L_b = 6$ are
accessible only if the collision energy exceeds the $E_c$ value of Eq.\
(\ref{eq:Ec}). It followed from Eqs.\ (\ref{eq:Rc}) and (\ref{eq:Ec}) that, if
$E_c$ increases, the critical field strength $B_c$ increases as well, and
higher partial waves can contribute over an increasingly wide range of fields.
This is also reflected in Figs.\ \ref{fig:pw_1muK} and \ref{fig:pw_1mK}. In the
ultracold regime, at a collision energy of 10$^{-6}$ K (Fig.\
\ref{fig:pw_1muK}), the $B_c$ values for $L_a=2\rightarrow L_b=4$ and $L_a=4\rightarrow L_b=6$
are relatively small and
the $s$-wave incident channel ($L_a=0 \rightarrow L_b=2$) is strongly dominant at all field
strengths above $B \approx 10^{-2}$ G.
At 10$^{-3}$ K, however, the $B_c$ values
for the higher partial-wave channels are much larger, and we find that
the $L_a=2$ and 4 incoming channels play a significant role at all magnetic field strengths
below $B \approx 10$ G.
A more detailed discussion on the energy dependence of the spin-inelastic cross section,
based on the Born approximation, will be given in the next section.

\subsection{Comparison with BA and DWBA}
\label{sec:results_B}
\begin{figure}
\centering
\includegraphics[width=15cm]{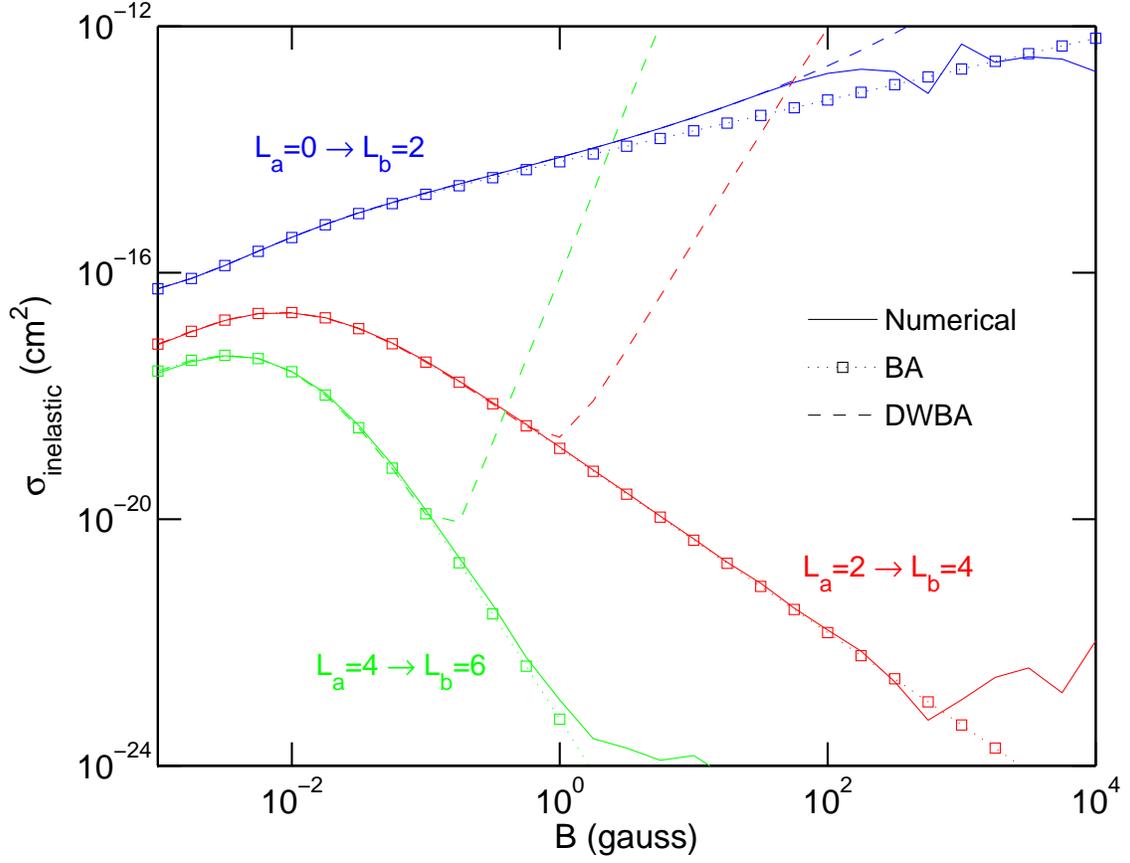}
\caption
  {\label{fig:pw_1muK_DWBA}
Total spin-inelastic collision cross sections for the $| M_{S_A} = 1, M_{S_B} = 1\rangle \rightarrow | M_{S_A} = 0, M_{S_B} = 0\rangle$ and
$| M_{S_A} = 1, M_{S_B} = 1\rangle \rightarrow | M_{S_A} = 1, M_{S_B} = 0\rangle + |M_{S_A} = 0, M_{S_B} = 1\rangle$
transitions of NH + NH, 
calculated as a function of magnetic field at 10$^{-6}$ K.
Different colors correspond to different $L_a \rightarrow L_b$ channel transitions. 
}
\end{figure}

\begin{figure}
\centering
\includegraphics[width=15cm]{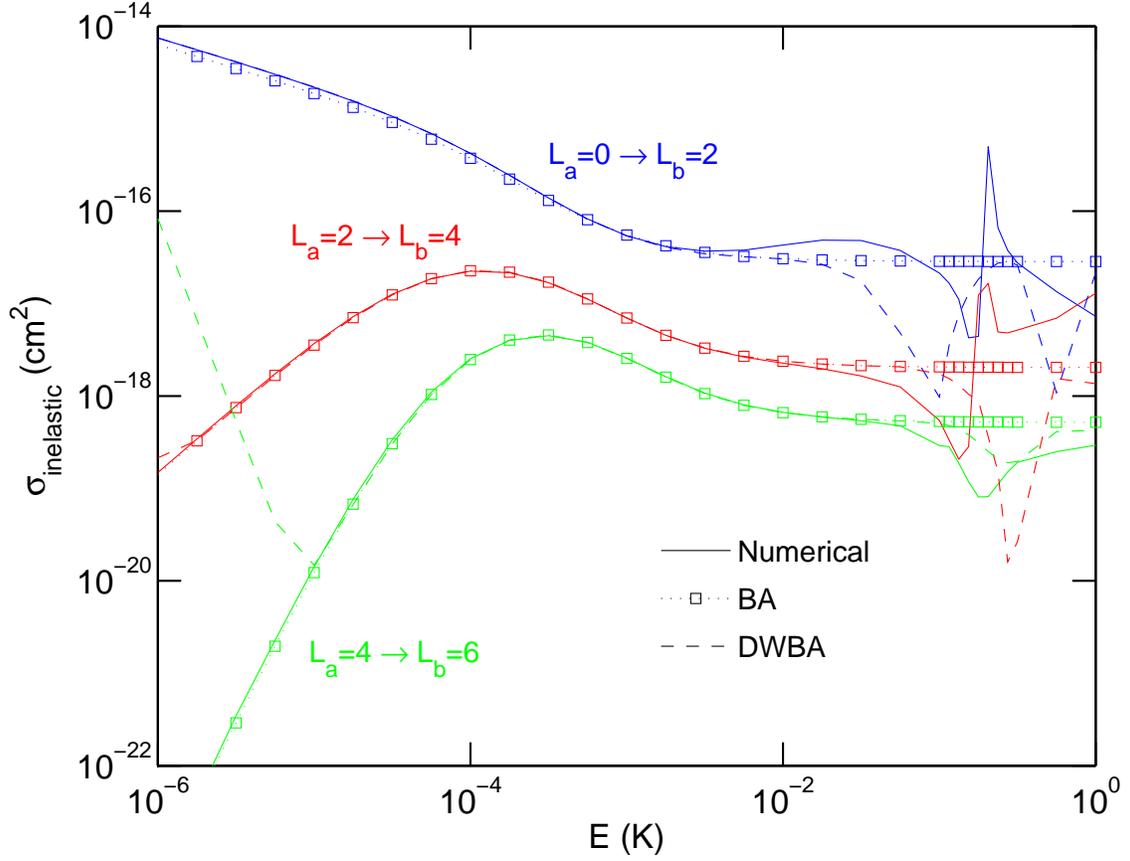}
\caption
  {\label{fig:pw_1G_DWBA}
Total spin-inelastic collision cross sections for the $| M_{S_A} = 1, M_{S_B} = 1\rangle \rightarrow | M_{S_A} = 0, M_{S_B} = 0\rangle$ and
$| M_{S_A} = 1, M_{S_B} = 1\rangle \rightarrow | M_{S_A} = 1, M_{S_B} = 0\rangle + |M_{S_A} = 0, M_{S_B} = 1\rangle$
transitions of NH + NH, 
calculated as a function of collision energy at a magnetic field strength of 1 G.
Different colors correspond to different $L_a \rightarrow L_b$ channel transitions. 
}
\end{figure}

Before comparing our numerical results with the analytical BA and DWBA
expressions, we must first point out that Eqs.\ (\ref{eq:sigBA}) and
(\ref{eq:sigDWBA}) apply only to collisions in which $M_{S_A}$ and $M_{S_B}$
each change at most by 1 and $L$ changes at most by 2.
Furthermore, the integral of Eq.\ (\ref{eq:intR_DWBA3}) is defined only if $L_b
> L_a+1$, and the DWBA cross section of Eq.\ (\ref{eq:sigDWBA}) is therefore
valid only for $L_a \rightarrow L_a+2$ transitions.  Figure
\ref{fig:pw_1muK_DWBA} shows the $|M^{(a)}_{S_A}=1, M^{(a)}_{S_B}=1, L_a
\rangle \rightarrow |M^{(b)}_{S_A}, M^{(b)}_{S_B}, L_a+2 \rangle$ cross
sections as a function of $B$ at a collision energy of 10$^{-6}$ K.
The cross sections are summed over all final states with $|M^{(b)}_{S_A}-M^{(a)}_{S_A}|\leq
1$ and $|M^{(b)}_{S_B}-M^{(a)}_{S_B}|\leq 1$. Figure \ref{fig:pw_1G_DWBA} shows the results
as a function of collision energy at a magnetic field strength of 1 G. It can
be seen that the BA results are in very good agreement with the cross sections
obtained from full CC calculations, in particular at low
magnetic fields and low collision energies. At high fields and energies, the numerical
cross sections exhibit several resonance features that arise mainly from the \textit{intramolecular}
spin-spin coupling term. Note that this coupling term is not included in the (DW)BA. 
Previous work has shown that the intramolecular spin-spin coupling becomes increasingly important as the
kinetic energy in the outgoing channel increases, and, for $B>10^2$ G and
$E>10^{-2}$ K, causes almost the same amount of spin relaxation as the intermolecular
magnetic dipolar interaction \cite{janssen:11a}. Hence, the BA result of Eq.\
(\ref{eq:sigBA}) deviates from the full CC result at high energies
and field strengths.

It can also be seen in Figs.\ \ref{fig:pw_1muK_DWBA} and \ref{fig:pw_1G_DWBA} that
the analytical distorted-wave BA cross section, which contains an extra term due to
the phase shift in the incoming channel, is in slightly better agreement with the
numerical $L_a=0 \rightarrow L_b=2$ cross section than the BA result. In particular,
Fig.\ \ref{fig:pw_1muK_DWBA} shows that the BA cross section for $L_a=0 \rightarrow L_b=2$ starts to deviate
from the CC calculations around $B\approx 1$ G, while the DWBA is accurate up
to $B \approx 100$ G. Thus, in the region between 1 and 100 G, the inelastic $L_a=0 \rightarrow L_b=2$ cross section
can be completely attributed to the intermolecular magnetic dipole interaction and to the phase shift in
the incident channel.
For the higher
partial-wave channels, however, the analytical DWBA cross section deviates significantly from
the full CC result at high fields and low energies. This is due to the
$(k_b/k_a)^{L_a+1}$ term in the expression for 
$\sigma^{\rm{DWBA}}_{a\rightarrow b}$ [Eq.\ (\ref{eq:sigDWBA})],
which tends to infinity if 
$k_a \ll k_b$. 
Even for very small phase shifts, this term will dominate
the DWBA inelastic cross section for $L_a > 0$ if the collision energy is small and the exothermicity is large.
More specifically, we estimate from Eq.\ (\ref{eq:sigDWBA}) that the DWBA cross section diverges
if $(k_a/k_b)^{2L_a+1} \approx T^{(M_a)}_{aa}$, and hence the effect is most pronounced for large $L_a$.
We point out that the origin of the $(k_b/k_a)^{L_a+1}$ term lies in the irregular spherical Bessel function
$y_{L_a}(k_a R)$ [Eq.\ (\ref{eq:intR_DWBA3})],
which enters the asymptotic wave function if the phase shift is nonzero.
At short range, the $y_{L_a}$ function tends to infinity and ultimately leads to the
unphysical behaviour observed in Figs.\ \ref{fig:pw_1muK_DWBA} and \ref{fig:pw_1G_DWBA}.
A possible remedy for this problem is to evaluate the integral of Eq.\ (\ref{eq:intR_DWBA3})
only for $R$-values larger than a certain cutoff radius. 
However, such an approach requires careful numerical analysis and falls out of the scope of the present study.
Nevetheless, based on the results shown in Figs.\ \ref{fig:pw_1muK_DWBA} and \ref{fig:pw_1G_DWBA}, we conclude that
the BA gives more meaningful results than the analytical DWBA at most of the energies and fields considered in this work.
As a final point, we note that the numerical phase shifts for the higher partial-wave channels are orders of magnitude smaller
than the $s$-wave scattering phase shift, and our DWBA results would not be substantially improved by
including a phase shift in the outgoing channel. 

As derived in Sec.\ \ref{sec:BA}, the threshold behaviour of the BA spin-inelastic cross section
in the limit of $k_a \ll k_b$ is given by $\sigma_{a\rightarrow b} \sim B^{1/2 - L_a}$
and $\sigma_{a\rightarrow b} \sim E^{L_a-1/2}$.
Indeed, we find that the $L_a=0\rightarrow L_b=2$, $L_a=2\rightarrow L_b=4$, and $L_a=4\rightarrow L_b=6$
inelastic cross sections at 10$^{-6}$ K behave as $B^{1/2}$, $B^{-3/2}$, and $B^{-7/2}$, respectively,
for field strengths above $B\approx 5\times10^{-2}$ G (see Fig.\ \ref{fig:pw_1muK_DWBA}).
Similarly, the cross sections at $B=1$ G follow an $E^{-1/2}$, $E^{3/2}$, and $E^{7/2}$ dependence,
respectively, at collision energies below $E \approx 10^{-4}$ K (see Fig.\ \ref{fig:pw_1G_DWBA}).
The validity regions of these threshold laws can also be explained in terms of the 
$V_{\rm{magn.dip}}$-induced avoided crossings discussed in Sec.\ \ref{sec:results_A}.
The critical magnetic field strengths below which the $L_a=2\rightarrow L_b=4$
and $L_a=4\rightarrow L_b=6$ crossings are energetically accessible are on the
order of $B_c \approx 10^{-2}$ G for a collision energy of 10$^{-6}$ K (see
Fig.\ \ref{fig:pw_1muK}). If the magnetic field strength exceeds $B_c$, the
crossings for the higher partial-wave channels are inaccessible and the corresponding 
scattering process 
can proceed only by (non-classical) tunneling through the centrifugal barrier.
Hence we find the quantum-mechanical threshold behaviour at fields
\textit{above} $B \approx 10^{-2}$ G.
For field strengths below $B_c$, the approximation of $k_a \ll k_b$ breaks down and 
the $B$-dependence follows from the explicit evaluation of Eq.\ (\ref{eq:sigBA}).
That is, the $B$-dependent threshold behaviour for $B < B_c$ is given by 
$(k_a/k_b)^{2L_a-1}$ multiplied by the hypergeometric function. If the magnetic field
is so small that $k_a \approx k_b$, the field dependence becomes negligible and the
cross section flattens off to a constant value. 
In order to explain the energy dependence, we apply Eqs.\ (\ref{eq:Rc}) and (\ref{eq:Ec})
to determine the lowest possible $E_c$ values at which the avoided curve crossings can occur.
At a magnetic field of 1 G, the corresponding values are $E_c = 5.8\times 10^{-5}$ K for the
$L_a=2\rightarrow L_b=4$ transition and $E_c = 1.2\times 10^{-4}$ K for $L_a=4\rightarrow L_b=6$.
Since the crossings for the higher partial-wave channels are inaccessible if $E < E_c$,
we recover the quantum-mechanical
threshold law at collision energies \textit{below} $E \approx 10^{-4}$ K. 

The results presented so far apply only to collisions where $M_{S_A}$ and $M_{S_B}$
decrease at most by 1 and $L$ increases by 2. The \textit{total} spin-inelastic cross section,
however, contains contributions from all (symmetry-allowed) outgoing partial waves
and all final states, i.e.\ also the states 
with $M_{S_A}^{(b)}=-1$ and $M_{S_B}^{(b)}=-1$. 
Let us now compare the BA and DWBA results with the numerical \textit{total}
spin-inelastic cross sections for magnetically trapped NH ($M_{S_A}^{(a)}=M_{S_B}^{(a)}=1$),
summed over all possible incoming partial waves and all outgoing channels.
The total BA cross section is obtained by performing the sums over $L_a$ and $L_b$ in Eq.\ (\ref{eq:sigBA})
for all possible (even) partial waves.
To calculate the total DWBA cross section, we perform the sums over $L_a$ and $L_b$ in Eq.\ (\ref{eq:sigDWBA})
for all possible (even) $L_a$ values and $L_b = L_a+2$.
Since the numerical scattering calculations were carried out for $L_{\rm{max}}=6$, we 
also took this maximum value for $L_a$ and $L_b$ in the (DW)BA expressions.

\begin{figure}
\centering
\includegraphics[width=15cm]{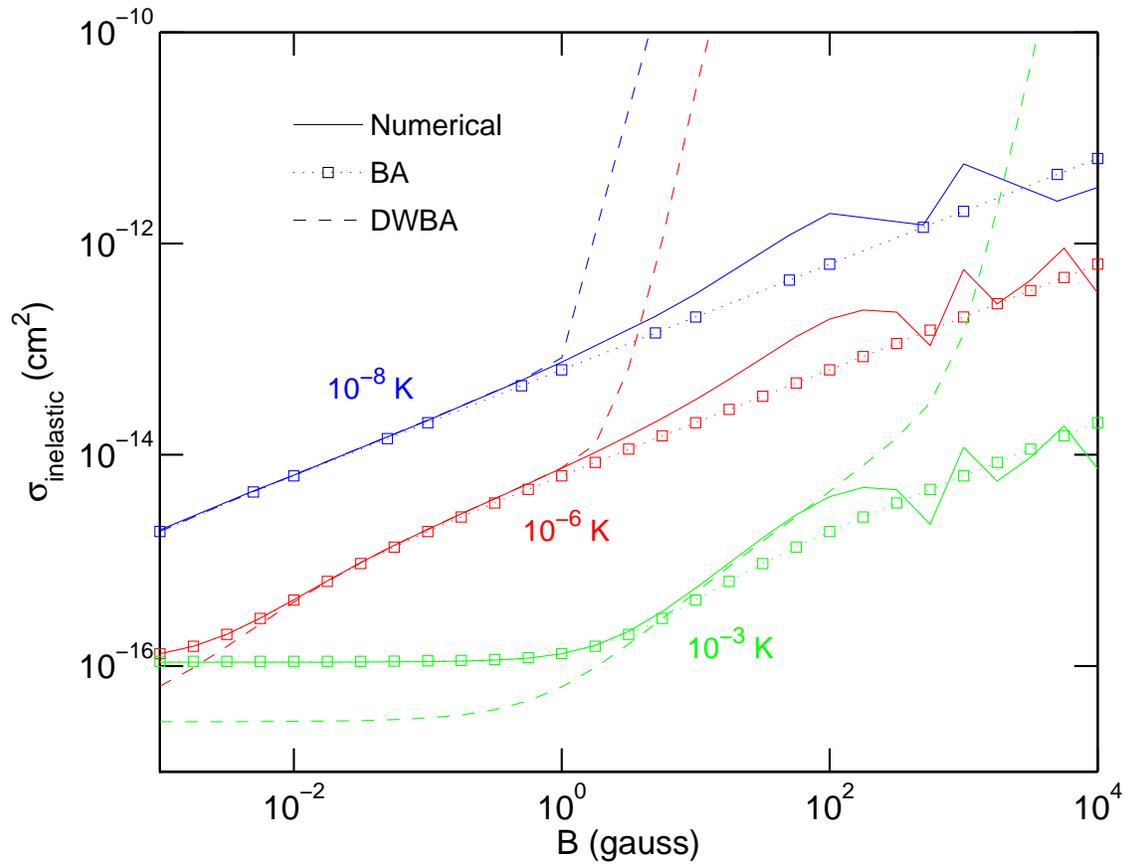}
\caption
  {\label{fig:scanB_totDWBA}
Total spin-inelastic collision cross sections for two magnetically trapped NH molecules, calculated
as a function of magnetic field. Different colors correspond to different collision energies.
}
\end{figure}

\begin{figure}
\centering
\includegraphics[width=15cm]{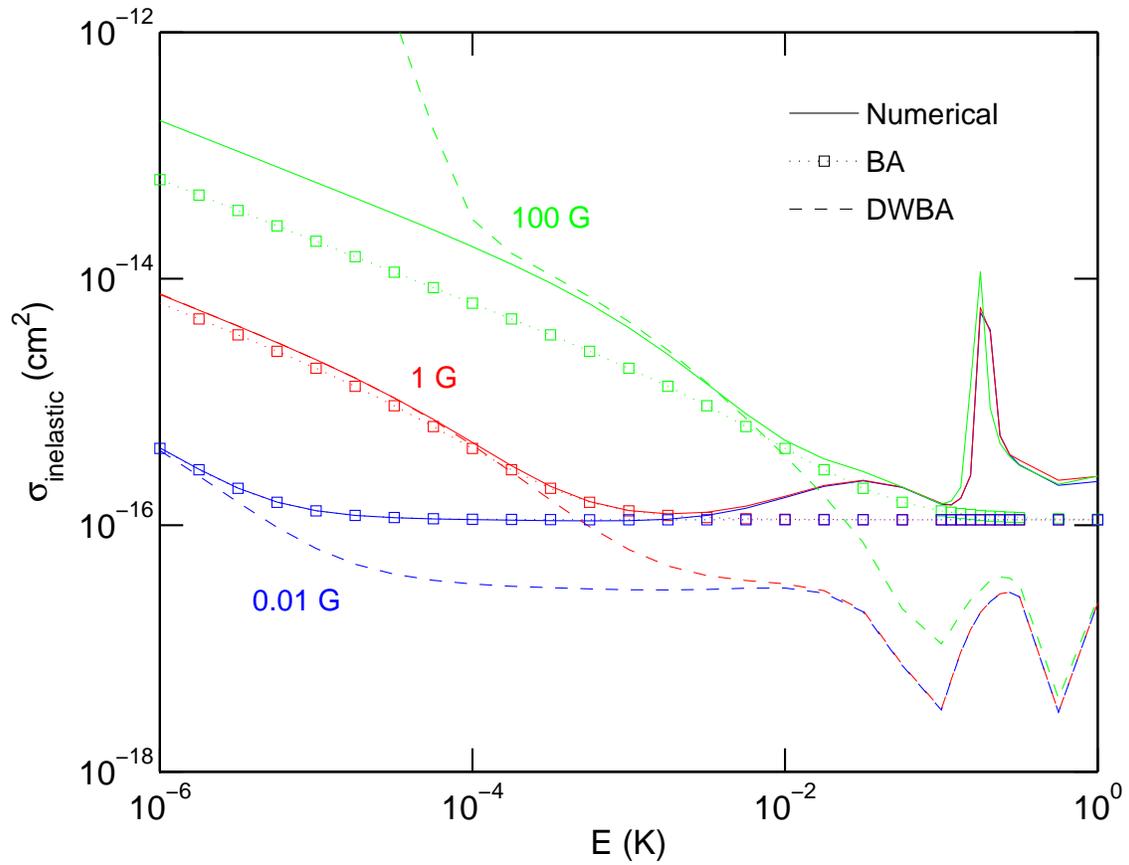}
\caption
  {\label{fig:scanE_totDWBA}
Total spin-inelastic collision cross sections for two magnetically trapped NH molecules, calculated
as a function of collision energy. Different colors correspond to different magnetic field strengths.
}
\end{figure}

The total inelastic cross sections are presented in Figs.\ \ref{fig:scanB_totDWBA} and
\ref{fig:scanE_totDWBA}.  It can be seen that the BA is generally in much
better agreement with the full CC result than the DWBA,
except for a small region near 10 G at 10$^{-3}$ K (Fig.\
\ref{fig:scanB_totDWBA}) and near 10$^{-3}$ K at 100 G (Fig.\
\ref{fig:scanE_totDWBA}), where $k_a \approx k_b$.
As noted previously, the deviation of the DWBA at
high $B$ and low $E$ is due to the $(k_b/k_a)^{L_a+1}$ term in Eq.\
(\ref{eq:sigDWBA}), which causes unphysical behaviour if $k_b \gg k_a$. At low
magnetic fields and relatively high energies, in particular at $E = 10^{-3}$ K
(see Fig.\ \ref{fig:scanB_totDWBA}), we find that the DWBA cross section also
deviates from the numerical result. In this region, the spin relaxation arises mainly
from the $L_a=2 \rightarrow L_b=2$ transition, and, to a smaller extent, also
from the $L_a=0 \rightarrow L_b=2$ and $L_a=2 \rightarrow L_b=0$ transitions.
Since the total DWBA cross section is restricted such that $L_b = L_a+2$, the
most dominant inelastic transitions at low $B$ are thus not included in the DWBA.
Similarly, the total DWBA cross section as a function of energy (Fig.\
\ref{fig:scanE_totDWBA}) shows a clear discrepancy with the numerical result at
$B = 0.01$ G for nearly all energies considered, and at $B = 1$ G for $E >
10^{-3}$ K. This is also due primarily to the missing $L_a=2 \rightarrow L_b=2$
channel transition. 

It can also be seen that the total BA cross section, which does include all
possible $L_a \rightarrow L_b$ transitions, agrees over a much wider range
of $E$ and $B$, but deviates from the full CC result at high fields and
high collision energies. As already
discussed in the first paragraph of this section, the deviation partly arises
from the \textit{intramolecular} spin-spin coupling term, which contributes
significantly to the numerical cross section as the kinetic energy in the
outgoing channel becomes large.  Moreover, the BA only includes contributions
from final states with $M_{S_A}=0,1$ and $M_{S_B}=0,1$, while the total
numerical cross section also contains terms with $M_{S_A}=-1$ and $M_{S_B}=-1$.
Previous work has shown that, as the intramolecular spin-spin coupling term
becomes increasingly important, the state-to-state cross sections for
$M_{S_A}^{(b)}=-1$ and/or $M_{S_B}^{(b)}=-1$ increase as well \cite{janssen:11a}.
Although
the dominant mechanism for $M_{S_i}^{(a)}=1 \rightarrow M_{S_i}^{(b)}=-1$
transitions is likely to be the intramolecular spin-spin term, which can decrease
$M_{S_i}$ by 2 directly in first order, the intermolecular magnetic dipolar
coupling term may also induce such spin-changing collisions in second order.
This effect is included only in the full CC calculation, and
hence this may represent another source of discrepancy between the BA and the
numerical result.

\section{Conclusions}
\label{sec:concl}
We have presented a detailed theoretical study on the role of the intermolecular
magnetic dipole-dipole interaction in (ultra)cold collisions of
magnetically trapped NH($X\,^3\Sigma^-$) molecules. The inelastic cross sections for Zeeman relaxation
have been obtained from rigorous coupled-channel calculations and from analytical results
based on the (distorted-wave) Born approximation. The derived expressions for the analytical
cross sections are valid for any two paramagnetic species for which the electronic spin and
its space-fixed projection are (approximately) good quantum numbers,
but we have applied them only to the NH + NH system.

We have found that the scattering of different partial waves, induced by the magnetic dipolar coupling, is
most easily understood by considering 
the adiabatic potential curves. The intermolecular
dipolar coupling term induces avoided crossings between certain adiabats at long range, which in turn
may lead to Zeeman relaxation. The cross-section behaviour as a function of energy
and magnetic field is, to a large extent, determined by whether or not these avoided
crossings are energetically accessible. Remarkably, the avoided crossings for higher partial
waves become lower in energy as the magnetic field strength decreases, implying that the
corresponding channels open up \textit{below} a certain critical field strength. Indeed, it
was found that the scattering of higher partial waves becomes increasingly important as the
exothermicity \textit{decreases}.

The validity regions of the analytical BA and DWBA have been determined
by comparison with numerical close-coupling calculations. We have found that the BA
is accurate over a wide range of collision energies and fields, but
starts to deviate from the numerical cross sections at energies above $\approx 10^{-2}$ K
and fields above $\approx 10^{2}$ G. This is mainly due to the effect of the \textit{intramolecular} spin-spin
coupling term, which is neglected in the BA but contributes significantly to the numerical cross section as the kinetic
energy in the outgoing channel becomes large.
The analytical distorted-wave Born approximation, which accounts for a phase shift
in the incident channel and thus represents a correction to the BA,
gives more accurate results than the BA in the case of $s$-wave scattering.
For higher partial-wave scattering, 
however,
and in particular at high magnetic fields and low energies, the DWBA
cross section shows unphysical behaviour and diverges to infinity. 
Furthermore, the derived DWBA expression is valid only for collisions where the partial-wave angular momentum is
increased by 2, while the total numerical cross section contains contributions from all possible
outgoing partial waves. More specifically, at fields below $\approx 1$ G and energies\
near 10$^{-3}$ K, the dominant contribution to the inelastic cross section is the $L_a =2 \rightarrow L_b=2$
channel transition, which is not included in the DWBA. Hence we conclude that the BA, which contains all
possible partial-wave contributions and does not show any unphysical behaviour, is generally in much better
agreement with the full CC result than the DWBA.

Although we have focused only on NH($^3\Sigma^-$) + NH($^3\Sigma^-$)
collisions in this study, the theory and main conclusions should be general for any two (ultra)cold
paramagnetic species.

\begin{acknowledgments}
We gratefully acknowledge Dr.\ Koos Gubbels for useful discussions and Dr.\ Piotr \.{Z}uchowski
for carefully reading the manuscript.
LMCJ and GCG thank the Council for
Chemical Sciences of the Netherlands Organization for Scientific Research
(CW-NWO) for financial support.
\end{acknowledgments}

\bibliography{vanderwaals}

\begin{thebibliography}{56}
\expandafter\ifx\csname natexlab\endcsname\relax\def\natexlab#1{#1}\fi
\expandafter\ifx\csname bibnamefont\endcsname\relax
  \def\bibnamefont#1{#1}\fi
\expandafter\ifx\csname bibfnamefont\endcsname\relax
  \def\bibfnamefont#1{#1}\fi
\expandafter\ifx\csname citenamefont\endcsname\relax
  \def\citenamefont#1{#1}\fi
\expandafter\ifx\csname url\endcsname\relax
  \def\url#1{\texttt{#1}}\fi
\expandafter\ifx\csname urlprefix\endcsname\relax\def\urlprefix{URL }\fi
\providecommand{\bibinfo}[2]{#2}
\providecommand{\eprint}[2][]{\url{#2}}

\bibitem[{\citenamefont{Greiner et~al.}(2002)\citenamefont{Greiner, Mandel,
  Esslinger, H\"{a}nsch, and Bloch}}]{greiner:02}
\bibinfo{author}{\bibfnamefont{M.}~\bibnamefont{Greiner}},
  \bibinfo{author}{\bibfnamefont{O.}~\bibnamefont{Mandel}},
  \bibinfo{author}{\bibfnamefont{T.}~\bibnamefont{Esslinger}},
  \bibinfo{author}{\bibfnamefont{T.~W.} \bibnamefont{H\"{a}nsch}},
  \bibnamefont{and} \bibinfo{author}{\bibfnamefont{I.}~\bibnamefont{Bloch}},
  \bibinfo{journal}{Nature} \textbf{\bibinfo{volume}{415}}, \bibinfo{pages}{39}
  (\bibinfo{year}{2002}).

\bibitem[{\citenamefont{Jaksch and Zoller}(2005)}]{jaksch:05}
\bibinfo{author}{\bibfnamefont{D.}~\bibnamefont{Jaksch}} \bibnamefont{and}
  \bibinfo{author}{\bibfnamefont{P.}~\bibnamefont{Zoller}},
  \bibinfo{journal}{Ann. Phys.} \textbf{\bibinfo{volume}{315}},
  \bibinfo{pages}{52} (\bibinfo{year}{2005}).

\bibitem[{\citenamefont{Micheli et~al.}(2006)\citenamefont{Micheli, Brennen,
  and Zoller}}]{micheli:06}
\bibinfo{author}{\bibfnamefont{A.}~\bibnamefont{Micheli}},
  \bibinfo{author}{\bibfnamefont{G.~K.} \bibnamefont{Brennen}},
  \bibnamefont{and} \bibinfo{author}{\bibfnamefont{P.}~\bibnamefont{Zoller}},
  \bibinfo{journal}{Nat. Phys.} \textbf{\bibinfo{volume}{2}},
  \bibinfo{pages}{341} (\bibinfo{year}{2006}).

\bibitem[{\citenamefont{Bloch et~al.}(2008)\citenamefont{Bloch, Dalibard, and
  Zwerger}}]{bloch:08}
\bibinfo{author}{\bibfnamefont{I.}~\bibnamefont{Bloch}},
  \bibinfo{author}{\bibfnamefont{J.}~\bibnamefont{Dalibard}}, \bibnamefont{and}
  \bibinfo{author}{\bibfnamefont{W.}~\bibnamefont{Zwerger}},
  \bibinfo{journal}{Rev. Mod. Phys.} \textbf{\bibinfo{volume}{80}},
  \bibinfo{pages}{885} (\bibinfo{year}{2008}).

\bibitem[{\citenamefont{DeMille}(2002)}]{demille:02}
\bibinfo{author}{\bibfnamefont{D.}~\bibnamefont{DeMille}},
  \bibinfo{journal}{Phys. Rev. Lett.} \textbf{\bibinfo{volume}{88}},
  \bibinfo{pages}{067901} (\bibinfo{year}{2002}).

\bibitem[{\citenamefont{Gulde et~al.}(2003)\citenamefont{Gulde, Riebe,
  Lancaster, Becher, Eschner, H\"{a}ffner, Schmidt-Kaler, Chuang, and
  Blatt}}]{gulde:03}
\bibinfo{author}{\bibfnamefont{S.}~\bibnamefont{Gulde}},
  \bibinfo{author}{\bibfnamefont{M.}~\bibnamefont{Riebe}},
  \bibinfo{author}{\bibfnamefont{G.~P.~T.} \bibnamefont{Lancaster}},
  \bibinfo{author}{\bibfnamefont{C.}~\bibnamefont{Becher}},
  \bibinfo{author}{\bibfnamefont{J.}~\bibnamefont{Eschner}},
  \bibinfo{author}{\bibfnamefont{H.}~\bibnamefont{H\"{a}ffner}},
  \bibinfo{author}{\bibfnamefont{F.}~\bibnamefont{Schmidt-Kaler}},
  \bibinfo{author}{\bibfnamefont{I.~L.} \bibnamefont{Chuang}},
  \bibnamefont{and} \bibinfo{author}{\bibfnamefont{R.}~\bibnamefont{Blatt}},
  \bibinfo{journal}{Nature} \textbf{\bibinfo{volume}{421}}, \bibinfo{pages}{48}
  (\bibinfo{year}{2003}).

\bibitem[{\citenamefont{Andr\'{e} et~al.}(2006)\citenamefont{Andr\'{e},
  DeMille, Doyle, Lukin, Maxwell, Rabl, Schoelkopf, and Zoller}}]{andre:06}
\bibinfo{author}{\bibfnamefont{A.}~\bibnamefont{Andr\'{e}}},
  \bibinfo{author}{\bibfnamefont{D.}~\bibnamefont{DeMille}},
  \bibinfo{author}{\bibfnamefont{J.~M.} \bibnamefont{Doyle}},
  \bibinfo{author}{\bibfnamefont{M.~D.} \bibnamefont{Lukin}},
  \bibinfo{author}{\bibfnamefont{S.~E.} \bibnamefont{Maxwell}},
  \bibinfo{author}{\bibfnamefont{P.}~\bibnamefont{Rabl}},
  \bibinfo{author}{\bibfnamefont{R.~J.} \bibnamefont{Schoelkopf}},
  \bibnamefont{and} \bibinfo{author}{\bibfnamefont{P.}~\bibnamefont{Zoller}},
  \bibinfo{journal}{Nat. Phys.} \textbf{\bibinfo{volume}{2}},
  \bibinfo{pages}{636} (\bibinfo{year}{2006}).

\bibitem[{\citenamefont{Lev et~al.}(2006)\citenamefont{Lev, Meyer, Hudson,
  Sawyer, Bohn, and Ye}}]{lev:06}
\bibinfo{author}{\bibfnamefont{B.~L.} \bibnamefont{Lev}},
  \bibinfo{author}{\bibfnamefont{E.~R.} \bibnamefont{Meyer}},
  \bibinfo{author}{\bibfnamefont{E.~R.} \bibnamefont{Hudson}},
  \bibinfo{author}{\bibfnamefont{B.~C.} \bibnamefont{Sawyer}},
  \bibinfo{author}{\bibfnamefont{J.~L.} \bibnamefont{Bohn}}, \bibnamefont{and}
  \bibinfo{author}{\bibfnamefont{J.}~\bibnamefont{Ye}}, \bibinfo{journal}{Phys.
  Rev. A} \textbf{\bibinfo{volume}{74}}, \bibinfo{pages}{061402}
  (\bibinfo{year}{2006}).

\bibitem[{\citenamefont{Fortier et~al.}(2007)\citenamefont{Fortier, Ashby,
  Bergquist, Delaney, Diddams, Heavner, Hollberg, Itano, Jefferts, Kim
  et~al.}}]{fortier:07}
\bibinfo{author}{\bibfnamefont{T.~M.} \bibnamefont{Fortier}},
  \bibinfo{author}{\bibfnamefont{N.}~\bibnamefont{Ashby}},
  \bibinfo{author}{\bibfnamefont{J.~C.} \bibnamefont{Bergquist}},
  \bibinfo{author}{\bibfnamefont{M.~J.} \bibnamefont{Delaney}},
  \bibinfo{author}{\bibfnamefont{S.~A.} \bibnamefont{Diddams}},
  \bibinfo{author}{\bibfnamefont{T.~P.} \bibnamefont{Heavner}},
  \bibinfo{author}{\bibfnamefont{L.}~\bibnamefont{Hollberg}},
  \bibinfo{author}{\bibfnamefont{W.~M.} \bibnamefont{Itano}},
  \bibinfo{author}{\bibfnamefont{S.~R.} \bibnamefont{Jefferts}},
  \bibinfo{author}{\bibfnamefont{K.}~\bibnamefont{Kim}}, \bibnamefont{et~al.},
  \bibinfo{journal}{Phys. Rev. Lett.} \textbf{\bibinfo{volume}{98}},
  \bibinfo{pages}{070801} (\bibinfo{year}{2007}).

\bibitem[{\citenamefont{Bethlem and Ubachs}(2009)}]{bethlem:09}
\bibinfo{author}{\bibfnamefont{H.~L.} \bibnamefont{Bethlem}} \bibnamefont{and}
  \bibinfo{author}{\bibfnamefont{W.}~\bibnamefont{Ubachs}},
  \bibinfo{journal}{Faraday Discuss.} \textbf{\bibinfo{volume}{142}},
  \bibinfo{pages}{25} (\bibinfo{year}{2009}).

\bibitem[{\citenamefont{Tarbutt et~al.}(2009)\citenamefont{Tarbutt, Hudson,
  Sauer, and Hinds}}]{tarbutt:09}
\bibinfo{author}{\bibfnamefont{M.~R.} \bibnamefont{Tarbutt}},
  \bibinfo{author}{\bibfnamefont{J.~J.} \bibnamefont{Hudson}},
  \bibinfo{author}{\bibfnamefont{B.~E.} \bibnamefont{Sauer}}, \bibnamefont{and}
  \bibinfo{author}{\bibfnamefont{E.~A.} \bibnamefont{Hinds}},
  \bibinfo{journal}{Faraday Discuss.} \textbf{\bibinfo{volume}{142}},
  \bibinfo{pages}{37} (\bibinfo{year}{2009}).

\bibitem[{\citenamefont{Poli et~al.}(2011)\citenamefont{Poli, Wang, Tarallo,
  Alberti, Prevedelli, and Tino}}]{poli:11}
\bibinfo{author}{\bibfnamefont{N.}~\bibnamefont{Poli}},
  \bibinfo{author}{\bibfnamefont{F.}~\bibnamefont{Wang}},
  \bibinfo{author}{\bibfnamefont{M.~G.} \bibnamefont{Tarallo}},
  \bibinfo{author}{\bibfnamefont{A.}~\bibnamefont{Alberti}},
  \bibinfo{author}{\bibfnamefont{M.}~\bibnamefont{Prevedelli}},
  \bibnamefont{and} \bibinfo{author}{\bibfnamefont{G.~M.} \bibnamefont{Tino}},
  \bibinfo{journal}{Phys. Rev. Lett.} \textbf{\bibinfo{volume}{106}},
  \bibinfo{pages}{038501} (\bibinfo{year}{2011}).

\bibitem[{\citenamefont{van~de Meerakker et~al.}(2005)\citenamefont{van~de
  Meerakker, Vanhaecke, van~der Loo, Groenenboom, and Meijer}}]{meerakker:05}
\bibinfo{author}{\bibfnamefont{S.~Y.~T.} \bibnamefont{van~de Meerakker}},
  \bibinfo{author}{\bibfnamefont{N.}~\bibnamefont{Vanhaecke}},
  \bibinfo{author}{\bibfnamefont{M.~P.~J.} \bibnamefont{van~der Loo}},
  \bibinfo{author}{\bibfnamefont{G.~C.} \bibnamefont{Groenenboom}},
  \bibnamefont{and} \bibinfo{author}{\bibfnamefont{G.}~\bibnamefont{Meijer}},
  \bibinfo{journal}{Phys. Rev. Lett.} \textbf{\bibinfo{volume}{95}},
  \bibinfo{pages}{013003} (\bibinfo{year}{2005}).

\bibitem[{\citenamefont{Gilijamse et~al.}(2006)\citenamefont{Gilijamse,
  Hoekstra, van~de Meerakker, Groenenboom, and Meijer}}]{gilijamse:06}
\bibinfo{author}{\bibfnamefont{J.~J.} \bibnamefont{Gilijamse}},
  \bibinfo{author}{\bibfnamefont{S.}~\bibnamefont{Hoekstra}},
  \bibinfo{author}{\bibfnamefont{S.~Y.~T.} \bibnamefont{van~de Meerakker}},
  \bibinfo{author}{\bibfnamefont{G.~C.} \bibnamefont{Groenenboom}},
  \bibnamefont{and} \bibinfo{author}{\bibfnamefont{G.}~\bibnamefont{Meijer}},
  \bibinfo{journal}{Science} \textbf{\bibinfo{volume}{313}},
  \bibinfo{pages}{1617} (\bibinfo{year}{2006}).

\bibitem[{\citenamefont{Gilijamse et~al.}(2007)\citenamefont{Gilijamse,
  Hoekstra, Meek, Mets\"al\"a, van~de Meerakker, Meijer, and
  Groenenboom}}]{gilijamse:07}
\bibinfo{author}{\bibfnamefont{J.~J.} \bibnamefont{Gilijamse}},
  \bibinfo{author}{\bibfnamefont{S.}~\bibnamefont{Hoekstra}},
  \bibinfo{author}{\bibfnamefont{S.~A.} \bibnamefont{Meek}},
  \bibinfo{author}{\bibfnamefont{M.}~\bibnamefont{Mets\"al\"a}},
  \bibinfo{author}{\bibfnamefont{S.~Y.~T.} \bibnamefont{van~de Meerakker}},
  \bibinfo{author}{\bibfnamefont{G.}~\bibnamefont{Meijer}}, \bibnamefont{and}
  \bibinfo{author}{\bibfnamefont{G.~C.} \bibnamefont{Groenenboom}},
  \bibinfo{journal}{J. Chem. Phys.} \textbf{\bibinfo{volume}{127}},
  \bibinfo{pages}{221102} (\bibinfo{year}{2007}).

\bibitem[{\citenamefont{Campbell et~al.}(2007)\citenamefont{Campbell, Tsikata,
  Lu, van Buuren, and Doyle}}]{campbell:07}
\bibinfo{author}{\bibfnamefont{W.~C.} \bibnamefont{Campbell}},
  \bibinfo{author}{\bibfnamefont{E.}~\bibnamefont{Tsikata}},
  \bibinfo{author}{\bibfnamefont{H.-I.} \bibnamefont{Lu}},
  \bibinfo{author}{\bibfnamefont{L.~D.} \bibnamefont{van Buuren}},
  \bibnamefont{and} \bibinfo{author}{\bibfnamefont{J.~M.} \bibnamefont{Doyle}},
  \bibinfo{journal}{Phys. Rev. Lett.} \textbf{\bibinfo{volume}{98}},
  \bibinfo{pages}{213001} (\bibinfo{year}{2007}).

\bibitem[{\citenamefont{Campbell et~al.}(2008)\citenamefont{Campbell,
  Groenenboom, Lu, Tsikata, and Doyle}}]{campbell:08}
\bibinfo{author}{\bibfnamefont{W.~C.} \bibnamefont{Campbell}},
  \bibinfo{author}{\bibfnamefont{G.~C.} \bibnamefont{Groenenboom}},
  \bibinfo{author}{\bibfnamefont{H.-I.} \bibnamefont{Lu}},
  \bibinfo{author}{\bibfnamefont{E.}~\bibnamefont{Tsikata}}, \bibnamefont{and}
  \bibinfo{author}{\bibfnamefont{J.~M.} \bibnamefont{Doyle}},
  \bibinfo{journal}{Phys. Rev. Lett.} \textbf{\bibinfo{volume}{100}},
  \bibinfo{pages}{083003} (\bibinfo{year}{2008}).

\bibitem[{\citenamefont{Sawyer et~al.}(2008)\citenamefont{Sawyer, Stuhl, Wang,
  Yeo, and Ye}}]{sawyer:08b}
\bibinfo{author}{\bibfnamefont{B.~C.} \bibnamefont{Sawyer}},
  \bibinfo{author}{\bibfnamefont{B.~K.} \bibnamefont{Stuhl}},
  \bibinfo{author}{\bibfnamefont{D.}~\bibnamefont{Wang}},
  \bibinfo{author}{\bibfnamefont{M.}~\bibnamefont{Yeo}}, \bibnamefont{and}
  \bibinfo{author}{\bibfnamefont{J.}~\bibnamefont{Ye}}, \bibinfo{journal}{Phys.
  Rev. Lett.} \textbf{\bibinfo{volume}{101}}, \bibinfo{pages}{203203}
  (\bibinfo{year}{2008}).

\bibitem[{\citenamefont{Krems}(2008)}]{krems:08}
\bibinfo{author}{\bibfnamefont{R.~V.} \bibnamefont{Krems}},
  \bibinfo{journal}{Phys. Chem. Chem. Phys.} \textbf{\bibinfo{volume}{10}},
  \bibinfo{pages}{4079} (\bibinfo{year}{2008}).

\bibitem[{\citenamefont{Campbell et~al.}(2009)\citenamefont{Campbell,
  Tscherbul, Lu, Tsikata, Krems, and Doyle}}]{campbell:09}
\bibinfo{author}{\bibfnamefont{W.~C.} \bibnamefont{Campbell}},
  \bibinfo{author}{\bibfnamefont{T.~V.} \bibnamefont{Tscherbul}},
  \bibinfo{author}{\bibfnamefont{H.-I.} \bibnamefont{Lu}},
  \bibinfo{author}{\bibfnamefont{E.}~\bibnamefont{Tsikata}},
  \bibinfo{author}{\bibfnamefont{R.~V.} \bibnamefont{Krems}}, \bibnamefont{and}
  \bibinfo{author}{\bibfnamefont{J.~M.} \bibnamefont{Doyle}},
  \bibinfo{journal}{Phys. Rev. Lett.} \textbf{\bibinfo{volume}{102}},
  \bibinfo{pages}{013003} (\bibinfo{year}{2009}).

\bibitem[{\citenamefont{Scharfenberg et~al.}(2010)\citenamefont{Scharfenberg,
  K{\l}os, Dagdigian, Alexander, Meijer, and van~de
  Meerakker}}]{scharfenberg:10}
\bibinfo{author}{\bibfnamefont{L.}~\bibnamefont{Scharfenberg}},
  \bibinfo{author}{\bibfnamefont{J.}~\bibnamefont{K{\l}os}},
  \bibinfo{author}{\bibfnamefont{P.~J.} \bibnamefont{Dagdigian}},
  \bibinfo{author}{\bibfnamefont{M.~H.} \bibnamefont{Alexander}},
  \bibinfo{author}{\bibfnamefont{G.}~\bibnamefont{Meijer}}, \bibnamefont{and}
  \bibinfo{author}{\bibfnamefont{S.~Y.~T.} \bibnamefont{van~de Meerakker}},
  \bibinfo{journal}{Phys. Chem. Chem. Phys.} \textbf{\bibinfo{volume}{12}},
  \bibinfo{pages}{10660} (\bibinfo{year}{2010}).

\bibitem[{\citenamefont{Ospelkaus et~al.}(2010)\citenamefont{Ospelkaus, Ni,
  Wang, de~Miranda, Neyenhuis, Qu\'em\'ener, Julienne, Bohn, Jin, and
  Ye}}]{ospelkaus:10}
\bibinfo{author}{\bibfnamefont{S.}~\bibnamefont{Ospelkaus}},
  \bibinfo{author}{\bibfnamefont{K.~K.} \bibnamefont{Ni}},
  \bibinfo{author}{\bibfnamefont{D.}~\bibnamefont{Wang}},
  \bibinfo{author}{\bibfnamefont{M.~H.~G.} \bibnamefont{de~Miranda}},
  \bibinfo{author}{\bibfnamefont{B.}~\bibnamefont{Neyenhuis}},
  \bibinfo{author}{\bibfnamefont{G.}~\bibnamefont{Qu\'em\'ener}},
  \bibinfo{author}{\bibfnamefont{P.~S.} \bibnamefont{Julienne}},
  \bibinfo{author}{\bibfnamefont{J.~L.} \bibnamefont{Bohn}},
  \bibinfo{author}{\bibfnamefont{D.~S.} \bibnamefont{Jin}}, \bibnamefont{and}
  \bibinfo{author}{\bibfnamefont{J.}~\bibnamefont{Ye}},
  \bibinfo{journal}{Science} \textbf{\bibinfo{volume}{327}},
  \bibinfo{pages}{853} (\bibinfo{year}{2010}).

\bibitem[{\citenamefont{Anderson et~al.}(1995)\citenamefont{Anderson, Ensher,
  Matthews, Wieman, and Cornell}}]{anderson:95a}
\bibinfo{author}{\bibfnamefont{M.~H.} \bibnamefont{Anderson}},
  \bibinfo{author}{\bibfnamefont{J.~R.} \bibnamefont{Ensher}},
  \bibinfo{author}{\bibfnamefont{M.~R.} \bibnamefont{Matthews}},
  \bibinfo{author}{\bibfnamefont{C.~E.} \bibnamefont{Wieman}},
  \bibnamefont{and} \bibinfo{author}{\bibfnamefont{E.~A.}
  \bibnamefont{Cornell}}, \bibinfo{journal}{Science}
  \textbf{\bibinfo{volume}{269}}, \bibinfo{pages}{198} (\bibinfo{year}{1995}).

\bibitem[{\citenamefont{Davis et~al.}(1995)\citenamefont{Davis, Mewes, Andrews,
  van Druten, Durfee, Kurn, and Ketterle}}]{davis:95}
\bibinfo{author}{\bibfnamefont{K.~B.} \bibnamefont{Davis}},
  \bibinfo{author}{\bibfnamefont{M.-O.} \bibnamefont{Mewes}},
  \bibinfo{author}{\bibfnamefont{M.~R.} \bibnamefont{Andrews}},
  \bibinfo{author}{\bibfnamefont{N.~J.} \bibnamefont{van Druten}},
  \bibinfo{author}{\bibfnamefont{D.~S.} \bibnamefont{Durfee}},
  \bibinfo{author}{\bibfnamefont{D.~M.} \bibnamefont{Kurn}}, \bibnamefont{and}
  \bibinfo{author}{\bibfnamefont{W.}~\bibnamefont{Ketterle}},
  \bibinfo{journal}{Phys. Rev. Lett.} \textbf{\bibinfo{volume}{75}},
  \bibinfo{pages}{3969} (\bibinfo{year}{1995}).

\bibitem[{\citenamefont{Shuman et~al.}(2010)\citenamefont{Shuman, Barry, and
  DeMille}}]{shuman:10}
\bibinfo{author}{\bibfnamefont{E.~S.} \bibnamefont{Shuman}},
  \bibinfo{author}{\bibfnamefont{J.~F.} \bibnamefont{Barry}}, \bibnamefont{and}
  \bibinfo{author}{\bibfnamefont{D.}~\bibnamefont{DeMille}},
  \bibinfo{journal}{Nature} \textbf{\bibinfo{volume}{467}},
  \bibinfo{pages}{820} (\bibinfo{year}{2010}).

\bibitem[{\citenamefont{Jones et~al.}(2006)\citenamefont{Jones, Tiesinga, Lett,
  and Julienne}}]{jones:06}
\bibinfo{author}{\bibfnamefont{K.~M.} \bibnamefont{Jones}},
  \bibinfo{author}{\bibfnamefont{E.}~\bibnamefont{Tiesinga}},
  \bibinfo{author}{\bibfnamefont{P.~D.} \bibnamefont{Lett}}, \bibnamefont{and}
  \bibinfo{author}{\bibfnamefont{P.~S.} \bibnamefont{Julienne}},
  \bibinfo{journal}{Rev. Mod. Phys.} \textbf{\bibinfo{volume}{78}},
  \bibinfo{pages}{483} (\bibinfo{year}{2006}).

\bibitem[{\citenamefont{K{\"{o}}hler et~al.}(2006)\citenamefont{K{\"{o}}hler,
  G\'{o}ral, and Julienne}}]{kohler:06}
\bibinfo{author}{\bibfnamefont{T.}~\bibnamefont{K{\"{o}}hler}},
  \bibinfo{author}{\bibfnamefont{K.}~\bibnamefont{G\'{o}ral}},
  \bibnamefont{and} \bibinfo{author}{\bibfnamefont{P.~S.}
  \bibnamefont{Julienne}}, \bibinfo{journal}{Rev. Mod. Phys.}
  \textbf{\bibinfo{volume}{78}}, \bibinfo{pages}{1311} (\bibinfo{year}{2006}).

\bibitem[{\citenamefont{Bethlem and Meijer}(2003)}]{bethlem:03}
\bibinfo{author}{\bibfnamefont{H.~L.} \bibnamefont{Bethlem}} \bibnamefont{and}
  \bibinfo{author}{\bibfnamefont{G.}~\bibnamefont{Meijer}},
  \bibinfo{journal}{Int. Rev. Phys. Chem.} \textbf{\bibinfo{volume}{22}},
  \bibinfo{pages}{73} (\bibinfo{year}{2003}).

\bibitem[{\citenamefont{Narevicius et~al.}(2008)\citenamefont{Narevicius,
  Libson, Parthey, Chavez, Narevicius, Even, and Raizen}}]{narevicius:08}
\bibinfo{author}{\bibfnamefont{E.}~\bibnamefont{Narevicius}},
  \bibinfo{author}{\bibfnamefont{A.}~\bibnamefont{Libson}},
  \bibinfo{author}{\bibfnamefont{C.~G.} \bibnamefont{Parthey}},
  \bibinfo{author}{\bibfnamefont{I.}~\bibnamefont{Chavez}},
  \bibinfo{author}{\bibfnamefont{J.}~\bibnamefont{Narevicius}},
  \bibinfo{author}{\bibfnamefont{U.}~\bibnamefont{Even}}, \bibnamefont{and}
  \bibinfo{author}{\bibfnamefont{M.~G.} \bibnamefont{Raizen}},
  \bibinfo{journal}{Phys. Rev. A} \textbf{\bibinfo{volume}{77}},
  \bibinfo{pages}{051401} (\bibinfo{year}{2008}).

\bibitem[{\citenamefont{Rieger et~al.}(2005)\citenamefont{Rieger, Junglen,
  Rangwala, Pinkse, and Rempe}}]{rieger:05}
\bibinfo{author}{\bibfnamefont{T.}~\bibnamefont{Rieger}},
  \bibinfo{author}{\bibfnamefont{T.}~\bibnamefont{Junglen}},
  \bibinfo{author}{\bibfnamefont{S.~A.} \bibnamefont{Rangwala}},
  \bibinfo{author}{\bibfnamefont{P.~W.~H.} \bibnamefont{Pinkse}},
  \bibnamefont{and} \bibinfo{author}{\bibfnamefont{G.}~\bibnamefont{Rempe}},
  \bibinfo{journal}{Phys. Rev. Lett.} \textbf{\bibinfo{volume}{95}},
  \bibinfo{pages}{173002} (\bibinfo{year}{2005}).

\bibitem[{\citenamefont{Weinstein et~al.}(1998)\citenamefont{Weinstein,
  deCarvalho, Guillet, Friedrich, and Doyle}}]{weinstein:98}
\bibinfo{author}{\bibfnamefont{J.~D.} \bibnamefont{Weinstein}},
  \bibinfo{author}{\bibfnamefont{R.}~\bibnamefont{deCarvalho}},
  \bibinfo{author}{\bibfnamefont{T.}~\bibnamefont{Guillet}},
  \bibinfo{author}{\bibfnamefont{B.}~\bibnamefont{Friedrich}},
  \bibnamefont{and} \bibinfo{author}{\bibfnamefont{J.~M.} \bibnamefont{Doyle}},
  \bibinfo{journal}{Nature} \textbf{\bibinfo{volume}{395}},
  \bibinfo{pages}{148} (\bibinfo{year}{1998}).

\bibitem[{\citenamefont{Sold\'{a}n et~al.}(2009)\citenamefont{Sold\'{a}n,
  \.{Z}uchowski, and Hutson}}]{soldan:09}
\bibinfo{author}{\bibfnamefont{P.}~\bibnamefont{Sold\'{a}n}},
  \bibinfo{author}{\bibfnamefont{P.~S.} \bibnamefont{\.{Z}uchowski}},
  \bibnamefont{and} \bibinfo{author}{\bibfnamefont{J.~M.}
  \bibnamefont{Hutson}}, \bibinfo{journal}{Faraday Discuss.}
  \textbf{\bibinfo{volume}{142}}, \bibinfo{pages}{191} (\bibinfo{year}{2009}).

\bibitem[{\citenamefont{Wallis and Hutson}(2009)}]{wallis:09}
\bibinfo{author}{\bibfnamefont{A.~O.~G.} \bibnamefont{Wallis}}
  \bibnamefont{and} \bibinfo{author}{\bibfnamefont{J.~M.}
  \bibnamefont{Hutson}}, \bibinfo{journal}{Phys. Rev. Lett.}
  \textbf{\bibinfo{volume}{103}}, \bibinfo{pages}{183201}
  (\bibinfo{year}{2009}).

\bibitem[{\citenamefont{Wallis et~al.}(2010)\citenamefont{Wallis, Longdon,
  \.{Z}uchowski, and Hutson}}]{wallis:10}
\bibinfo{author}{\bibfnamefont{A.~O.~G.} \bibnamefont{Wallis}},
  \bibinfo{author}{\bibfnamefont{E.~J.~J.} \bibnamefont{Longdon}},
  \bibinfo{author}{\bibfnamefont{P.~S.} \bibnamefont{\.{Z}uchowski}},
  \bibnamefont{and} \bibinfo{author}{\bibfnamefont{J.~M.} \bibnamefont{Hutson}}
  (\bibinfo{year}{2010}), \eprint{arXiv:1009.5505}.

\bibitem[{\citenamefont{Barletta et~al.}(2009)\citenamefont{Barletta, Tennyson,
  and Barker}}]{barletta:09}
\bibinfo{author}{\bibfnamefont{P.}~\bibnamefont{Barletta}},
  \bibinfo{author}{\bibfnamefont{J.}~\bibnamefont{Tennyson}}, \bibnamefont{and}
  \bibinfo{author}{\bibfnamefont{P.~F.} \bibnamefont{Barker}},
  \bibinfo{journal}{New J. Phys.} \textbf{\bibinfo{volume}{11}},
  \bibinfo{pages}{055029} (\bibinfo{year}{2009}).

\bibitem[{\citenamefont{Barletta et~al.}(2010)\citenamefont{Barletta, Tennyson,
  and Barker}}]{barletta:10}
\bibinfo{author}{\bibfnamefont{P.}~\bibnamefont{Barletta}},
  \bibinfo{author}{\bibfnamefont{J.}~\bibnamefont{Tennyson}}, \bibnamefont{and}
  \bibinfo{author}{\bibfnamefont{P.~F.} \bibnamefont{Barker}},
  \bibinfo{journal}{New J. Phys.} \textbf{\bibinfo{volume}{12}},
  \bibinfo{pages}{113002} (\bibinfo{year}{2010}).

\bibitem[{\citenamefont{Avdeenkov and Bohn}(2001)}]{avdeenkov:01}
\bibinfo{author}{\bibfnamefont{A.~V.} \bibnamefont{Avdeenkov}}
  \bibnamefont{and} \bibinfo{author}{\bibfnamefont{J.~L.} \bibnamefont{Bohn}},
  \bibinfo{journal}{Phys. Rev. A} \textbf{\bibinfo{volume}{64}},
  \bibinfo{pages}{052703} (\bibinfo{year}{2001}).

\bibitem[{\citenamefont{Janssen
  et~al.}(2011{\natexlab{a}})\citenamefont{Janssen, \.{Z}uchowski, van~der
  Avoird, Hutson, and Groenenboom}}]{janssen:11}
\bibinfo{author}{\bibfnamefont{L.~M.~C.} \bibnamefont{Janssen}},
  \bibinfo{author}{\bibfnamefont{P.~S.} \bibnamefont{\.{Z}uchowski}},
  \bibinfo{author}{\bibfnamefont{A.}~\bibnamefont{van~der Avoird}},
  \bibinfo{author}{\bibfnamefont{J.~M.} \bibnamefont{Hutson}},
  \bibnamefont{and} \bibinfo{author}{\bibfnamefont{G.~C.}
  \bibnamefont{Groenenboom}}, \bibinfo{journal}{J. Chem. Phys.}
  \textbf{\bibinfo{volume}{134}}, \bibinfo{pages}{accepted}
  (\bibinfo{year}{2011}{\natexlab{a}}).

\bibitem[{\citenamefont{Janssen
  et~al.}(2011{\natexlab{b}})\citenamefont{Janssen, \.{Z}uchowski, van~der
  Avoird, Groenenboom, and Hutson}}]{janssen:11a}
\bibinfo{author}{\bibfnamefont{L.~M.~C.} \bibnamefont{Janssen}},
  \bibinfo{author}{\bibfnamefont{P.~S.} \bibnamefont{\.{Z}uchowski}},
  \bibinfo{author}{\bibfnamefont{A.}~\bibnamefont{van~der Avoird}},
  \bibinfo{author}{\bibfnamefont{G.~C.} \bibnamefont{Groenenboom}},
  \bibnamefont{and} \bibinfo{author}{\bibfnamefont{J.~M.}
  \bibnamefont{Hutson}}, \bibinfo{journal}{Phys. Rev. A}
  \textbf{\bibinfo{volume}{83}}, \bibinfo{pages}{022713}
  (\bibinfo{year}{2011}{\natexlab{b}}).

\bibitem[{\citenamefont{Ketterle and van Druten}(1996)}]{ketterle:96}
\bibinfo{author}{\bibfnamefont{W.}~\bibnamefont{Ketterle}} \bibnamefont{and}
  \bibinfo{author}{\bibfnamefont{N.~J.} \bibnamefont{van Druten}},
  \bibinfo{journal}{Adv. Atom. Mol. Opt. Phys.} \textbf{\bibinfo{volume}{37}},
  \bibinfo{pages}{181} (\bibinfo{year}{1996}).

\bibitem[{\citenamefont{Gerton et~al.}(1999)\citenamefont{Gerton, Sackett,
  Frew, and Hulet}}]{gerton:99}
\bibinfo{author}{\bibfnamefont{J.~M.} \bibnamefont{Gerton}},
  \bibinfo{author}{\bibfnamefont{C.~A.} \bibnamefont{Sackett}},
  \bibinfo{author}{\bibfnamefont{B.~J.} \bibnamefont{Frew}}, \bibnamefont{and}
  \bibinfo{author}{\bibfnamefont{R.~G.} \bibnamefont{Hulet}},
  \bibinfo{journal}{Phys. Rev. A} \textbf{\bibinfo{volume}{59}},
  \bibinfo{pages}{1514} (\bibinfo{year}{1999}).

\bibitem[{\citenamefont{Tscherbul et~al.}(2010)\citenamefont{Tscherbul,
  K{\l}os, Dalgarno, Zygelman, Pavlovic, Hummon, Lu, Tsikata, and
  Doyle}}]{tscherbul:10}
\bibinfo{author}{\bibfnamefont{T.~V.} \bibnamefont{Tscherbul}},
  \bibinfo{author}{\bibfnamefont{J.}~\bibnamefont{K{\l}os}},
  \bibinfo{author}{\bibfnamefont{A.}~\bibnamefont{Dalgarno}},
  \bibinfo{author}{\bibfnamefont{B.}~\bibnamefont{Zygelman}},
  \bibinfo{author}{\bibfnamefont{Z.}~\bibnamefont{Pavlovic}},
  \bibinfo{author}{\bibfnamefont{M.~T.} \bibnamefont{Hummon}},
  \bibinfo{author}{\bibfnamefont{H.-I.} \bibnamefont{Lu}},
  \bibinfo{author}{\bibfnamefont{E.}~\bibnamefont{Tsikata}}, \bibnamefont{and}
  \bibinfo{author}{\bibfnamefont{J.~M.} \bibnamefont{Doyle}},
  \bibinfo{journal}{Phys. Rev. A} \textbf{\bibinfo{volume}{82}},
  \bibinfo{pages}{042718} (\bibinfo{year}{2010}).

\bibitem[{\citenamefont{Hensler et~al.}(2003)\citenamefont{Hensler, Werner,
  Griesmaier, Schmidt, G\"{o}rlitz, Pfau, Giovanazzi, and
  Rza\.{z}ewski}}]{hensler:03}
\bibinfo{author}{\bibfnamefont{S.}~\bibnamefont{Hensler}},
  \bibinfo{author}{\bibfnamefont{J.}~\bibnamefont{Werner}},
  \bibinfo{author}{\bibfnamefont{A.}~\bibnamefont{Griesmaier}},
  \bibinfo{author}{\bibfnamefont{P.~O.} \bibnamefont{Schmidt}},
  \bibinfo{author}{\bibfnamefont{A.}~\bibnamefont{G\"{o}rlitz}},
  \bibinfo{author}{\bibfnamefont{T.}~\bibnamefont{Pfau}},
  \bibinfo{author}{\bibfnamefont{S.}~\bibnamefont{Giovanazzi}},
  \bibnamefont{and}
  \bibinfo{author}{\bibfnamefont{K.}~\bibnamefont{Rza\.{z}ewski}},
  \bibinfo{journal}{Appl. Phys. B} \textbf{\bibinfo{volume}{77}},
  \bibinfo{pages}{765} (\bibinfo{year}{2003}).

\bibitem[{\citenamefont{Hummon et~al.}(2011)\citenamefont{Hummon, Tscherbul,
  K{\l}os, Lu, Tsikata, Campbell, Dalgarno, and Doyle}}]{hummon:11}
\bibinfo{author}{\bibfnamefont{M.~T.} \bibnamefont{Hummon}},
  \bibinfo{author}{\bibfnamefont{T.~V.} \bibnamefont{Tscherbul}},
  \bibinfo{author}{\bibfnamefont{J.}~\bibnamefont{K{\l}os}},
  \bibinfo{author}{\bibfnamefont{H.-I.} \bibnamefont{Lu}},
  \bibinfo{author}{\bibfnamefont{E.}~\bibnamefont{Tsikata}},
  \bibinfo{author}{\bibfnamefont{W.~C.} \bibnamefont{Campbell}},
  \bibinfo{author}{\bibfnamefont{A.}~\bibnamefont{Dalgarno}}, \bibnamefont{and}
  \bibinfo{author}{\bibfnamefont{J.~M.} \bibnamefont{Doyle}},
  \bibinfo{journal}{Phys. Rev. Lett.} \textbf{\bibinfo{volume}{106}},
  \bibinfo{pages}{053201} (\bibinfo{year}{2011}).

\bibitem[{\citenamefont{Krems and Dalgarno}(2004)}]{krems:04b}
\bibinfo{author}{\bibfnamefont{R.~V.} \bibnamefont{Krems}} \bibnamefont{and}
  \bibinfo{author}{\bibfnamefont{A.}~\bibnamefont{Dalgarno}},
  \bibinfo{journal}{J. Chem. Phys.} \textbf{\bibinfo{volume}{120}},
  \bibinfo{pages}{2296} (\bibinfo{year}{2004}).

\bibitem[{\citenamefont{Hutson and Green}()}]{molscat:94}
\bibinfo{author}{\bibfnamefont{J.~M.} \bibnamefont{Hutson}} \bibnamefont{and}
  \bibinfo{author}{\bibfnamefont{S.}~\bibnamefont{Green}}, \bibinfo{note}{{\sc
  molscat} computer code, version 14 (1994), distributed by Collaborative
  Computational Project No. 6 of the Engineering and Physical Sciences Research
  Council (UK)}.

\bibitem[{\citenamefont{Gonz\'{a}lez-Mart\'{i}nez and
  Hutson}(2007)}]{gonzalez:07}
\bibinfo{author}{\bibfnamefont{M.~L.} \bibnamefont{Gonz\'{a}lez-Mart\'{i}nez}}
  \bibnamefont{and} \bibinfo{author}{\bibfnamefont{J.~M.}
  \bibnamefont{Hutson}}, \bibinfo{journal}{Phys. Rev. A}
  \textbf{\bibinfo{volume}{75}}, \bibinfo{pages}{022702}
  (\bibinfo{year}{2007}).

\bibitem[{\citenamefont{Alexander and Manolopoulos}(1987)}]{alexander:87}
\bibinfo{author}{\bibfnamefont{M.~H.} \bibnamefont{Alexander}}
  \bibnamefont{and} \bibinfo{author}{\bibfnamefont{D.~E.}
  \bibnamefont{Manolopoulos}}, \bibinfo{journal}{J. Chem. Phys.}
  \textbf{\bibinfo{volume}{86}}, \bibinfo{pages}{2044} (\bibinfo{year}{1987}).

\bibitem[{\citenamefont{Moerdijk and Verhaar}(1996)}]{moerdijk:96}
\bibinfo{author}{\bibfnamefont{A.~J.} \bibnamefont{Moerdijk}} \bibnamefont{and}
  \bibinfo{author}{\bibfnamefont{B.~J.} \bibnamefont{Verhaar}},
  \bibinfo{journal}{Phys. Rev. A} \textbf{\bibinfo{volume}{53}},
  \bibinfo{pages}{R19} (\bibinfo{year}{1996}).

\bibitem[{\citenamefont{Avdeenkov and Bohn}(2005)}]{avdeenkov:05}
\bibinfo{author}{\bibfnamefont{A.~V.} \bibnamefont{Avdeenkov}}
  \bibnamefont{and} \bibinfo{author}{\bibfnamefont{J.~L.} \bibnamefont{Bohn}},
  \bibinfo{journal}{Phys. Rev. A} \textbf{\bibinfo{volume}{71}},
  \bibinfo{pages}{022706} (\bibinfo{year}{2005}).

\bibitem[{\citenamefont{Kajita}(2006)}]{kajita:06}
\bibinfo{author}{\bibfnamefont{M.}~\bibnamefont{Kajita}},
  \bibinfo{journal}{Phys. Rev. A} \textbf{\bibinfo{volume}{74}},
  \bibinfo{pages}{032710} (\bibinfo{year}{2006}).

\bibitem[{\citenamefont{Zygelman}(2010)}]{zygelman:10}
\bibinfo{author}{\bibfnamefont{B.}~\bibnamefont{Zygelman}},
  \bibinfo{journal}{Phys. Rev. A} \textbf{\bibinfo{volume}{81}},
  \bibinfo{pages}{032506} (\bibinfo{year}{2010}).

\bibitem[{\citenamefont{Messiah}(1969)}]{messiah:69}
\bibinfo{author}{\bibfnamefont{A.}~\bibnamefont{Messiah}},
  \emph{\bibinfo{title}{Quantum Mechanics}} (\bibinfo{publisher}{North
  Holland}, \bibinfo{address}{Amsterdam}, \bibinfo{year}{1969}).

\bibitem[{\citenamefont{Abramowitz and Stegun}(1964)}]{abramowitz:64}
\bibinfo{author}{\bibfnamefont{M.}~\bibnamefont{Abramowitz}} \bibnamefont{and}
  \bibinfo{author}{\bibfnamefont{I.~A.} \bibnamefont{Stegun}},
  \emph{\bibinfo{title}{Handbook of Mathematical Functions}}
  (\bibinfo{publisher}{National Bureau of Standards},
  \bibinfo{address}{Washington, D.C.}, \bibinfo{year}{1964}),
  \urlprefix\url{http://www.math.sfu.ca/~cbm/aands}.

\bibitem[{\citenamefont{Volpi and Bohn}(2002)}]{volpi:02}
\bibinfo{author}{\bibfnamefont{A.}~\bibnamefont{Volpi}} \bibnamefont{and}
  \bibinfo{author}{\bibfnamefont{J.~L.} \bibnamefont{Bohn}},
  \bibinfo{journal}{Phys. Rev. A} \textbf{\bibinfo{volume}{65}},
  \bibinfo{pages}{052712} (\bibinfo{year}{2002}).

\bibitem[{\citenamefont{Tscherbul et~al.}(2009)\citenamefont{Tscherbul,
  Suleimanov, Aquilanti, and Krems}}]{tscherbul:09c}
\bibinfo{author}{\bibfnamefont{T.~V.} \bibnamefont{Tscherbul}},
  \bibinfo{author}{\bibfnamefont{Y.~V.} \bibnamefont{Suleimanov}},
  \bibinfo{author}{\bibfnamefont{V.}~\bibnamefont{Aquilanti}},
  \bibnamefont{and} \bibinfo{author}{\bibfnamefont{R.~V.} \bibnamefont{Krems}},
  \bibinfo{journal}{New J. Phys.} \textbf{\bibinfo{volume}{11}},
  \bibinfo{pages}{055021} (\bibinfo{year}{2009}).

\end{thebibliography}

\clearpage

\clearpage

\clearpage

\clearpage

\clearpage

\clearpage

\end{document}